\documentclass[journal,11pt,onecolumn,draftclsnofoot]{IEEEtran}
\usepackage{algorithmic}
\usepackage{algorithm}
\usepackage{array}
\usepackage{caption,subcaption}
\usepackage{textcomp}
\usepackage{stfloats}
\usepackage[hidelinks]{hyperref}
\usepackage{url}
\usepackage{verbatim}
\usepackage{graphicx}
\usepackage{orcidlink}
\usepackage{lipsum}
\usepackage{setspace}
\usepackage{outlines}
\usepackage{soul}
\usepackage{amsmath,amsfonts,amsthm,amssymb}
\usepackage{scalefnt}
\usepackage[acronym,shortcuts,nonumberlist]{glossaries}
\usepackage[bottom]{footmisc}
\usepackage{multirow}
\usepackage{stackengine}
\usepackage{outlines}
\usepackage{tikz}
\usepackage{cite}
\usepackage{colortbl}
\usepackage{rotating}

\newcommand{\herm}{^{\mathsf{H}}}

\makeatletter
\DeclareRobustCommand{\iscircle}{\mathord{\mathpalette\is@circle\relax}}
\newcommand\is@circle[2]{%
\begingroup
\sbox\z@{\raisebox{\depth}{$\m@th#1\bigcirc$}}%
\sbox\tw@{$#1\square$}%
\resizebox{!}{\ht\tw@}{\usebox{\z@}}%
\endgroup
}
\makeatother
\definecolor{teagreen}{rgb}{0.82, 0.94, 0.75}
\definecolor{flavescent}{rgb}{0.97, 0.91, 0.56}
\definecolor{indianred}{rgb}{0.8, 0.36, 0.36}
\newcommand\good{\cellcolor{teagreen}}
\newcommand\medium{\cellcolor{flavescent}}
\newcommand\bad{\cellcolor{indianred}}

\newacronym{OFDM}{OFDM}{orthogonal frequency division multiplexing}
\newacronym{DFT-s-OFDM}{DFT-s-OFDM}{DFT-spread OFDM}
\newacronym{FBMC}{FBMC}{filter bank multi-carrier}
\newacronym{GFDM}{GFDM}{generalized frequency division multiplexing}
\newacronym{OCDM}{OCDM}{orthogonal chirp division multiplexing}
\newacronym{AFDM}{AFDM}{affine frequency division multiplexing}
\newacronym{OTFS}{OTFS}{orthogonal time frequency space}
\newacronym{T-OTFS}{T-OTFS}{transcendentally-rotated OTFS}
\newacronym{ODDM}{ODDM}{orthogonal delay-Doppler division multiplexing}
\newacronym{SWHM}{SWHM}{sparse Walsh-Hadamard multiplexing}
\newacronym{FMCW}{FMCW}{frequency modulated continuous wave}
\newacronym{OTSM}{OTSM}{orthogonal time sequency modulation}

\newacronym{LCT}{LCT}{linear canonical transform}
\newacronym{ZT}{ZT}{Zak transform}
\newacronym{DZT}{DZT}{discrete Zak transform}
\newacronym{IDZT}{IDZT}{inverse discrete Zak transform}
\newacronym{FT}{FT}{Fourier transform}
\newacronym{IFT}{IFT}{inverse Fourier transform}
\newacronym{DFT}{DFT}{discrete Fourier transform}
\newacronym{IDFT}{IDFT}{inverse discrete Fourier transform}
\newacronym{AFT}{AFT}{affine Fourier transform}
\newacronym{DAFT}{DAFT}{discrete affine Fourier transform}
\newacronym{IDAFT}{IDAFT}{inverse discrete affine Fourier transform}
\newacronym{SFFT}{SFFT}{symplectic finite Fourier transform}
\newacronym{ISFFT}{ISFFT}{inverse symplectic finite Fourier transform}
\newacronym{HT}{HT}{Heisenberg transform}
\newacronym{SISO}{SISO}{single-input single-output}
\newacronym{MIMO}{MIMO}{multiple-input multiple-output}
\newacronym{WT}{WT}{Wigner transform}
\newacronym{frFT}{frFT}{fractional Fourier transform}
\newacronym{IfrFT}{IfrFT}{inverse fractional Fourier transform}
\newacronym{fnT}{fnT}{Fresnel transform}
\newacronym{IfnT}{IfnT}{inverse Fresnel transform}
\newacronym{LT}{LT}{Laplace transform}
\newacronym{ILT}{ILT}{inverse Laplace transform}
\newacronym{ISAC}{ISAC}{integrated sensing and communications}
\newacronym{JCAS}{JCAS}{joint communications and sensing}
\newacronym{EM}{EM}{electromagnetic}
\newacronym{CP}{CP}{cyclic prefix}
\newacronym{B5G}{B5G}{beyond fifth generation}
\newacronym{3GPP}{3GPP}{$3^{\rm{rd}}$ generation partnership project}           
\newacronym{4G}{4G}{fourth generation}                                          
\newacronym{5G}{5G}{fifth generation}                                           
\newacronym{6G}{6G}{sixth generation}
\newacronym{LTV}{LTV}{linear time-variant}
\newacronym{LTI}{LTI}{linear time-invariant}
\newacronym{LTVM}{LTVM}{linear time-variant multipath}
\newacronym{TV}{TV}{time-variant}
\newacronym{TI}{TI}{time-invariant}
\newacronym{1D}{1D}{one-dimensional}
\newacronym{2D}{2D}{two-dimensional}
\newacronym{3D}{3D}{three-dimensional}
\newacronym{NTN}{NTN}{non-terrestrial network}
\newacronym{LEO}{LEO}{low earth orbit}
\newacronym{IoT}{IoT}{Internet-of-Things}
\newacronym{mmWave}{mmWave}{millimeter-wave}
\newacronym{THz}{THz}{Terahertz}
\newacronym{V2X}{V2X}{vehicle-to-everything}
\newacronym{RCC}{RCC}{radar-communication coexistence}
\newacronym{C-V2X}{C-V2X}{Cellular-V2X}                                             
\newacronym{NR}{NR}{New Radio}                                                      
\newacronym{ETSI}{ETSI}{European Telecommunications Standards Institute}            
\newacronym{EHF}{EHF}{extremely high-frequency}
\newacronym{BER}{BER}{bit-error-rate}
\newacronym{ICI}{ICI}{inter-carrier interference}
\newacronym{SotA}{SotA}{state-of-the-art}
\newacronym{DoF}{DoF}{degrees-of-freedom}
\newacronym{UAV}{UAV}{unmanned aerial vehicle}
\newacronym{UE}{UE}{user equipment}
\newacronym{AP}{AP}{access point}
\newacronym{CIR}{CIR}{channel impulse response}
\newacronym{CSI}{CSI}{channel state information}
\newacronym{SNR}{SNR}{signal-to-noise ratio}
\newacronym{AWGN}{AWGN}{additive white Gaussian noise}
\newacronym{ML}{ML}{maximum likelihood}
\newacronym{OOB}{OOB}{out-of-band}
\newacronym{MX}{MX}{multiplexing}
\newacronym{DMX}{DMX}{demultiplexing}
\newacronym{AoA}{AoA}{angle-of-arrival}
\newacronym{AoD}{AoD}{angle-of-departure}
\newacronym{5GAA}{5GAA}{5G Automotive Association}                                  
\newacronym{6G-IA}{6G-IA}{6G Smart Networks and Services Industry Association}      
\newacronym{MUSIC}{MUSIC}{MUltiple SIgnal Classification}
\newacronym{ESPRIT}{ESPRIT}{estimation of signal parameters via rotational invariance techniques}
\newacronym{DP}{DP}{detection problem}
\newacronym{EP}{EP}{estimation problem}
\newacronym{KPI}{KPI}{key performance indicator}
\newacronym{ISI}{ISI}{inter-symbol interference}
\newacronym{TVIRF}{TVIRF}{time-variant impulse response function}
\newacronym{TVTF}{TVTF}{time-variant transfer function}
\newacronym{DVIRF}{DVIRF}{Doppler-variant impulse response function}
\newacronym{DVTF}{DVTF}{Doppler-variant transfer function}
\newacronym{DDSF}{DDSF}{delay-Doppler spread function}
\newacronym{PAPR}{PAPR}{peak-to-average power ratio}
\newacronym{RCS}{RCS}{radar cross-section}
\newacronym{PIR}{PIR}{peak interference residual}
\newacronym{PSR}{PSR}{peak-to-sidelobe ratio}
\newacronym{PA}{PA}{power amplifier}
\newacronym{RF}{RF}{radio frequency}
\newacronym{JCDE}{JCDE}{joint channel and data estimation}
\begin{document}

%
\title{ $~$ \\[-3ex] {\normalsize{\textbf{\color{red}Please find the official IEEE \underline{published} version of this article on IEEE Xplore \href{https://ieeexplore.ieee.org/abstract/document/10769778}{{\color{blue}[here]}}} \textbf{\color{red}and cite as:} \\[-1ex]} 

\color{cyan} H. S. Rou \emph{et al.}, ``From Orthogonal Time-Frequency Space to Affine Frequency-Division Multiplexing:  \\[-1.5ex] 

A comparative study of next-generation waveforms for integrated sensing and communications in dou-\\[-1.5ex]

bly dispersive channels [Special Issue on Signal Processing for the Integrated Sensing and Commun- \\[-6ex]

ications Revolution]," \emph{in IEEE Signal Processing Magazine}, vol. 41, no. 5, pp. 71-86, Sept. 2024}\\[0.5ex]

From OTFS to AFDM:\! A Comparative Study \\[-0.5ex] of Next-Generation Waveforms for ISAC \\[-0.5ex] in Doubly-Dispersive Channels}

\author{
\vspace{1ex}
Hyeon Seok Rou,~\IEEEmembership{Graduate Student Member,~IEEE},\\
Giuseppe Thadeu Freitas de Abreu,~\IEEEmembership{Senior Member,~IEEE},
Junil Choi,~\IEEEmembership{Senior Member,~IEEE}, \\
David Gonz{\'{a}}lez G.,~\IEEEmembership{Senior Member,~IEEE},
Marios Kountouris,~\IEEEmembership{Fellow,~IEEE}, \\
Yong Liang Guan,~\IEEEmembership{Senior Member,~IEEE}, and
Osvaldo Gonsa. 

\vspace{-4ex}

}

\markboth{IEEE Signal Processing Magazine Special Issue: ``Signal Processing for the ISAC Revolution"}%
{Rou \MakeLowercase{\textit{et al.}}:}



\maketitle

%
\begin{abstract}
\label{sec:abstract}

Next-generation wireless systems will offer \ac{ISAC} functionalities not only in order to enable new applications, but also as a means to mitigate challenges such as doubly-dispersive channels, which arise in high mobility scenarios and/or at \ac{mmWave} and \ac{THz} bands.
An emerging approach to accomplish these goals is the design of new waveforms, which draw from the inherent relationship between the doubly-dispersive nature of \ac{TV} channels and the environmental features of scatterers manifested in the form of multipath delays and Doppler shifts.
Examples of such waveforms are the delay-Doppler domain \ac{OTFS} and the recently proposed chirp domain \ac{AFDM}, both of which seek to simultaneously combat the detrimental effects of double selectivity and exploit them for the estimation (or sensing) of environmental information.
This article aims to provide a consolidated and comprehensive overview of the signal processing techniques required to support reliable \ac{ISAC} over doubly-dispersive channels in \ac{B5G}/\ac{6G} systems, with an emphasis on \ac{OTFS} and \ac{AFDM} waveforms, as those, together with the traditional \ac{OFDM} waveform, suffice to elaborate on the most relevant properties of the trend.
The analysis shows that \ac{OTFS} and \ac{AFDM} indeed enable significantly improved robustness against \ac{ICI} arising from Doppler shifts compared to \ac{OFDM}.
In addition, the inherent delay-Doppler domain orthogonality of the \ac{OTFS} and \ac{AFDM} effective channels is found to provide significant advantages for the design and the performance of integrated sensing functionalities.

\end{abstract}

\glsresetall

\section{Introduction}
\label{sec:introduction}
\vspace{-0.5ex}

It is expected that \ac{B5G} and \ac{6G} wireless systems will employ \ac{EHF} technologies, operating in the \ac{mmWave} and \ac{THz} bands \cite{Rappaport_Access19} as a means to support applications \cite{Vo_MNA22}, such as \Ac{IoT}, edge computing and smart cities; and scenarios such as \ac{V2X}, high-speed rail, and \acp{NTN}, which are often subjected to heterogeneous and high-mobility conditions \cite{ZhongIEEENet2022}.
High-mobility scenarios are known to pose a significant challenge to wireless communications systems due to the resulting doubly-dispersive wireless channel, also referred to as \ac{TV} multipath, or time-frequency selectivity \cite{Bliss_13}.
Such heterogeneous scattering environments deteriorate the received signal in the form of path delays and Doppler shifts, resulting in \ac{ISI} and \ac{ICI} which can drastically decrease communication performance under conventional and highly effective modulation schemes, such as \ac{OFDM} \cite{Wang_TWC06}.

Concomitant with this challenge, there is a growing expectation that \ac{B5G} and \ac{6G} systems will offer \ac{ISAC} capabilities, possibly with unified hardware and signal processing techniques \cite{Zhang_CST22}.
In addition to providing environment perception and accurate/reliable localization information to serve the aforementioned applications, the enhancements introduced by \ac{ISAC} are fundamental to improve spectrum and energy efficiency, and to lower hardware costs of systems operating in high-mobility scenarios \cite{ZhongIEEENet2022}.

While it is difficult to foresee which of the upcoming generations/standards -- \textit{i.e.,} \ac{B5G} or \ac{6G} -- will see \ac{ISAC} adopted and implemented into commercial systems, the topic is one of the most intensively discussed among pre-standardization fora on wireless systems in recent years\footnote{See \href{https://5gaa.org/}{\tt https://5gaa.org/} and \href{https://6g-ia.eu/}{\tt https://6g-ia.eu/} for additional information on \acs{5GAA} and \acs{6G-IA}, respectively.}, with notable examples being the \ac{6G-IA}, where \ac{ISAC} has been identified as a priority technology for its members, and the \acf{5GAA}, where \ac{ISAC} is considered an enabling technology for \ac{C-V2X} services.
Although it is reasonable to anticipate that any form of practically deployed \acs{5G}-based \ac{ISAC} will likely leverage \ac{OFDM}, more specifically \ac{CP}-\ac{OFDM} and \ac{DFT-s-OFDM} for downlink/sidelink, and uplink, respectively, but for \ac{6G}, new waveforms, such as \ac{OTFS} and \ac{AFDM}, should be considered to fully enable the potential of \ac{ISAC}, as will be investigated in this article.
In fact, important standardization bodies such as \ac{ETSI} and the \ac{3GPP}, which produce technical specifications for mobile broadband systems worldwide, have recently added \ac{ISAC} to their work plans and roadmaps, with \ac{ETSI} launching a new group dedicated to \ac{ISAC} in November 2023.


In line with this trend, novel waveforms have been recently proposed which, thanks to their ability to retain symbol orthogonality under doubly-dispersive conditions, are both robust to high-mobility and advantageous for \ac{ISAC}, as they inherently enable the estimation of environmental parameters, such as distance and velocity of scatterer objects (\textit{i.e.,} delay and Doppler shifts).
One of the most popular methods is \ac{OTFS} signaling \cite{Wei_WC21}, which leverages the \ac{ISFFT} in order to modulate a \ac{2D} grid of information symbols directly in the delay-Doppler domain, gaining great attention for high-mobility \ac{B5G} systems thanks to its superior performance compared to currently used waveforms such as \ac{OFDM} \cite{Anwar_WCNC20}.

It is easy to show, indeed, that the full delay-Doppler representation of the channel in \ac{OTFS} inherently conveys the velocity and range information of the scatterers in the form of the respective multipath delays and Doppler shifts, thus implying significant benefits in terms of \ac{ISAC}.
As a consequence, a plethora of \ac{OTFS}-based \ac{ISAC} techniques have been proposed to extract the delay and Doppler parameters of the resolvable paths directly from the \ac{CSI}, which have been shown to compete with the sensing performances of \ac{OFDM} and \ac{FMCW} radars, with higher robustness to mobility and achievable capacity \cite{Gaudio_TWC20}.

An alternative strategy to design \ac{ISAC}-friendly and mobility-robust waveforms is to employ chirp-based multicarrier approaches \cite{Ouyang_TC16}.
While the chirp-domain design is attractive due to the inherent spread-spectrum property and potential for full-duplex operations, an important and common drawback of these earlier approaches is the lack of adaptability to the channel delay and Doppler spreads, which is a consequence of the non-parametrizable transforms in their design.

A more recent take on the idea, which seeks to mitigate the latter drawback, is the \ac{AFDM} waveform \cite{Bemani_TWC23}, which leverages the \ac{IDAFT} \cite{Healy_LCT15} in order to modulate information symbols into a \textit{twisted} time-frequency domain, yielding the desired delay-Doppler orthogonality while maintaining the necessary flexibility.
The optimizable parametrization of \ac{AFDM} is further accompanied by other desirable properties, such as full diversity guarantee and increased throughput \cite{Bemani_TWC23}, making \ac{AFDM} a strong candidate of \ac{ISAC}-enabling waveform for \ac{B5G} and \ac{6G} systems.


This article aims to offer a thorough analysis of the fundamentals and the future of \ac{ISAC} technology in heterogeneous high-mobility scenarios, in the form of a comprehensive comparison of prominent candidate waveforms, focusing on \ac{OTFS} and \ac{AFDM}.
The analysis reveals that the novel delay-Doppler orthogonal designs of \ac{OTFS} and \ac{AFDM} benefit the signal processing for both communication and sensing functionalities, advocating the integration of the two. 
These insights may hold significant interest and value not only for academia, but also for standardization engineers across various industry verticals who are increasingly participating in the development of future generations of mobile broadband systems.

The remainder of the article is organized as follows:
the fundamental system models and the required \ac{ISAC} signal processing techniques for the doubly-dispersive wireless channel\footnote{The importance of modelling the doubly-dispersive channel, especially for \ac{ISAC} applications, is a highly relevant problem currently discussed in both academia and standardization \cite{RP234069}.}, emphasizing the inherent transformations between time, frequency, delay, and Doppler dimensions, are described in Section~\ref{sec:sigproc_fundamentals}.
In Section~\ref{sec:waveforms}, the signal models of the identified candidate waveforms for \ac{B5G}/\ac{6G} \ac{ISAC} in doubly-dispersive environments are consolidated, highlighting their interrelationships in terms of the multiplexing domain, transmitter structure, and the core \textit{\ac{LCT}}. 
In Section~\ref{sec:radar_sensing}, we discuss the radar sensing techniques leveraging the identified candidate waveforms in terms of the radar target \ac{DP} and radar parameter \ac{EP}, elaborating on signal processing techniques and solutions categorized into correlation-based methods, and direct/indirect \ac{CSI}-based approaches.
In Section~\ref{sec:comparative_analysis}, the candidate waveforms are compared with basis on different \acp{KPI} for both communications and radar sensing performances, in addition to implications onto hardware implementation, requirements, and potential challenges.  
Finally, the key insights provided by the article are summarized, and some future directions of the research are identified.

\vspace{-0.25ex}
\section{Signal Processing Fundamentals of Doubly-Dispersive Channels}
\label{sec:sigproc_fundamentals}

A long history of research on wireless communications has resulted in the identification and characterization of two fundamental and distinct types of small-scale fading effects, namely frequency- and time-selectivity, also known as time and frequency dispersion, respectively.
In particular, an \ac{EM} signal propagated through a given path is subject to a specific \textit{path delay} proportional to the total propagation distance between the transmitter and the receiver, and a \textit{Doppler shift}\footnote{We remark that in the related literature, and therefore also in this article, the term \textit{Doppler shift} is often used in a broad sense, including spectral shifts of the propagated signal resulting from phenomena other than the actual Doppler effect, such as frequency offsets and low-frequency phase noise at the local oscillators.} proportional to the relative velocities among transmitter, receiver, and scatterer, and the carrier frequency.

In a channel with multiple distinguishable propagation paths, the different copies of the originally transmitted signal with varying time delays and Doppler shifts are superposed at the receiver, resulting in interference that impacts on the reliability and performance of the wireless communication link, unless appropriate signal processing techniques are employed.
In this section, we first consolidate the fundamental doubly-dispersive channel model with all of its representations in the time, frequency, delay, and Doppler domains, along with the associated transformation methods, followed by the corresponding signal processing mechanisms available to process the received signal, by the efficient representation of the input-output relationship leveraging a circular convolution matrix.

\subsection{The Doubly-Dispersive Channel Model}
\label{subsec:dd_channel}
Consider a wireless channel between a transmitter and receiver   which is modelled via $P$ significant propagation paths, where each $p$-th resolvable path, with $p \in \{1,\cdots,P\}$, is respectively described by a corresponding complex fading coefficient $h_p \in \mathbb{C}$, path delay $\tau_p \in [0, \tau^\mathrm{max}]$, and Doppler shift $\nu_p \in [-\nu^\mathrm{max}, +\nu^\mathrm{max}]$.
The corresponding delay and Doppler spreads of such a doubly-dispersive channel are characterized by  the maximum delay $\tau^\mathrm{max}$ [s] and the maximum Doppler shift $\pm \nu^\mathrm{max}$ [Hz], such that the channel can be described by the \ac{LTV}\footnote{The term ``\acf{LTV} system'' is not to be confused with linear systems with \textit{time-varying delays} -- \textit{i.e.,} systems with delay drifts, where $\tau_p(t)$) -- which are also commonly described as \ac{LTV}. In this article, we only consider time-invariant delays, in compliance with the related literature on doubly-dispersive channels, e.g., \cite{Bliss_13, Wei_WC21, Bemani_TWC23}.} relationship between the input and the output signals.
The \ac{LTV} channel is most commonly represented as a \ac{TVIRF} in the time-delay domain, given by
\begin{equation}
h(t, \tau) \triangleq \sum_{p=1}^{P} h_p \cdot e^{j2\pi \nu_p t} \cdot \delta(\tau - \tau_p),
\label{eq:TVIR}
\vspace{-1ex}
\end{equation}  
where $j \triangleq \sqrt{-1}$ is the elementary imaginary number, $t$ and $\tau$ denote the instantaneous time and path delay, respectively, and $\delta(x)$ is the unit impulse function defined by $\delta(x) = 1 ~\mathrm{iff}~ x = 0$.

Alternatively, the \ac{TVIRF} in the time-delay domain can also be represented in other domains by leveraging appropriate linear transforms \cite{Healy_LCT15}.
For example, the representation in the time-frequency domain is known as \ac{TVTF}, which is obtained by a \ac{FT} on the \ac{TVIRF} over the delay domain, \textit{i.e.,}
\vspace{-0.5ex}
\begin{equation}
H(t, f) \triangleq \underset{\tau \!\rightarrow \!f}{\mathcal{F}}[ h(t, \tau) ]
= \int_{-\infty}^{+\infty} h(t, \tau) \cdot e^{-j2\pi\tau f} \;{\rm d}\tau
= \sum_{p=1}^{P} h_p \cdot e^{j2\pi \nu_p t} \cdot  e^{-j2\pi \tau_p f},
\label{eq:TVTF}
\vspace{-1ex}
\end{equation}
where $f$ is the instantaneous frequency and $\mathcal{F}[\; \cdot \;]$ denotes the continuous \ac{FT} operator.

\begin{figure}[t]
\centering
\subfloat[\acs{TVTF} in the time-frequency domain.]{\includegraphics[width=0.495\textwidth]{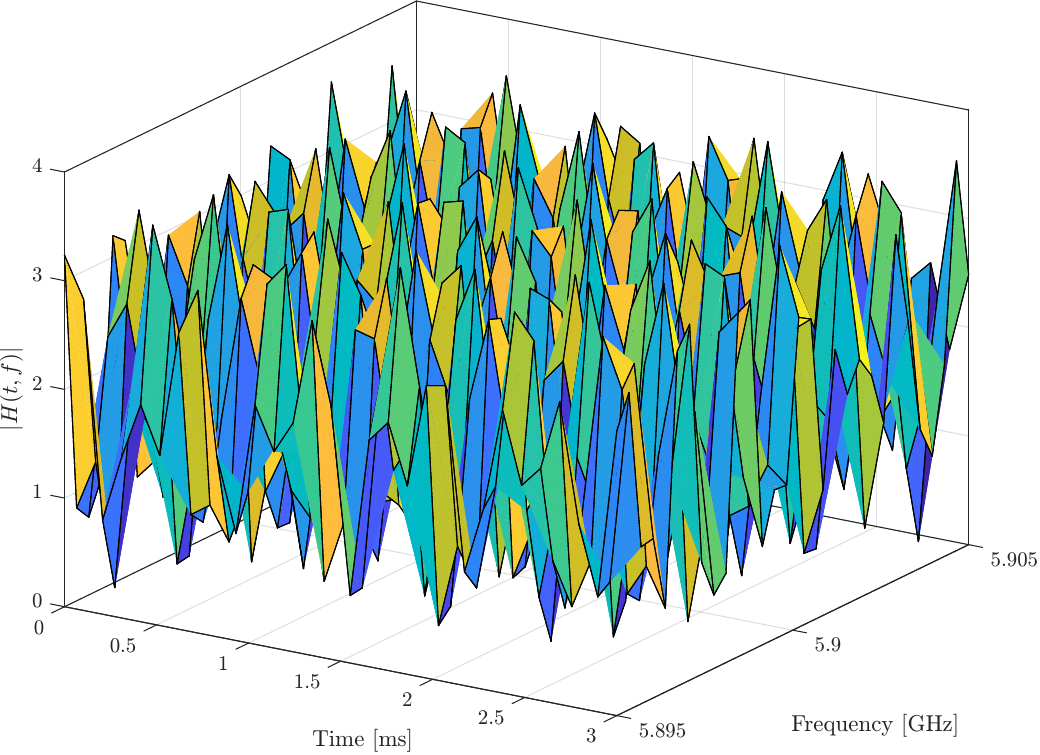}%
\label{fig:TVTF}}
\hfil
\subfloat[\acs{DVIRF} in the delay-Doppler domain.]{\includegraphics[width=0.495\textwidth]{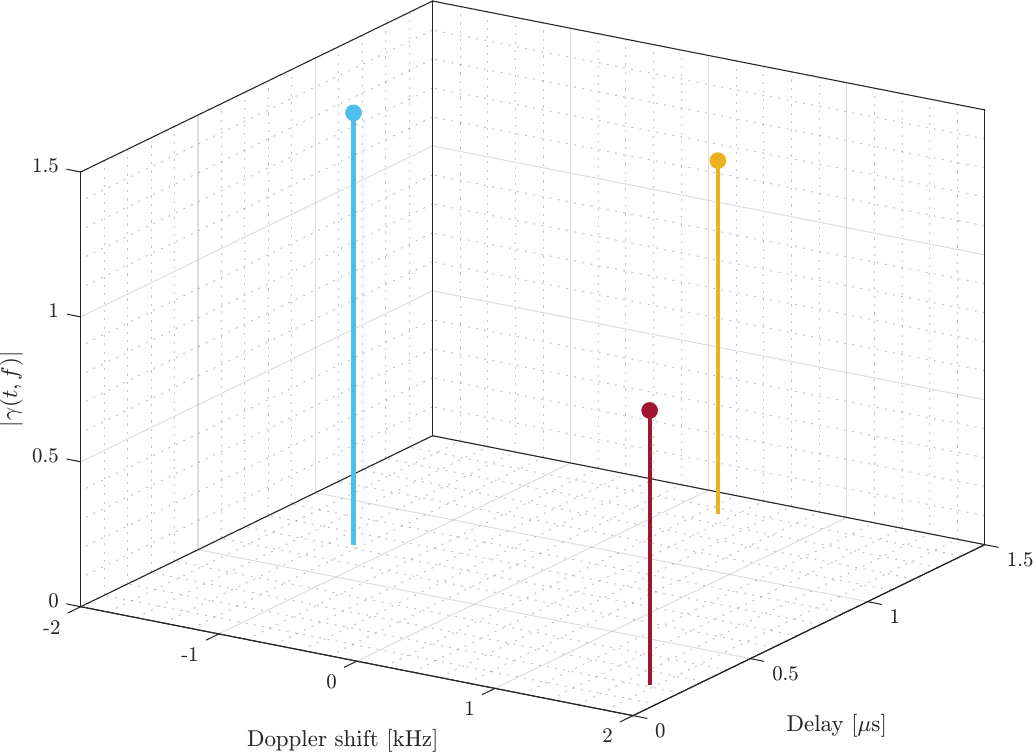}%
\label{fig:DVIRF}}
\caption{The doubly-dispersive channel representations with $P = 3$ resolvable paths, with carrier frequency of $5.9\,$GHz and signal bandwidth of $10\,$MHz (following the IEEE 802.11p vehicular environment specifications).
The different paths are illustrated by unique colors in Fig. \ref{fig:DVIRF}.}
\label{fig:LTV_representations}
\vspace{-4ex}
\end{figure}

The \ac{TVTF} in the time-frequency domain readily highlights both time and frequency dispersion effects of the channel, visible in the two fast-varying exponential terms dependent on the instantaneous time $t$ and instantaneous frequency $f$, respectively at a rate of the Doppler frequency $\nu_p$ and delay $\tau_p$ of the corresponding $p$-th propagation path, as illustrated in Fig. \ref{fig:TVTF}.
Conversely, the \ac{DVIRF} in the delay-Doppler domain is obtained by an \ac{FT} on the \ac{TVIRF} over the time domain, that is
\vspace{-0.5ex}
\begin{equation}
\gamma(\nu, \tau) \triangleq \underset{t \rightarrow \nu}{\mathcal{F}}[ h(t, \tau) ] =
\int_{-\infty}^{+\infty} h(t, \tau) \cdot e^{-j2\pi \nu t} \;{\rm d}t = 
\sum_{p=1}^{P} h_p \cdot \delta(\nu - \nu_p) \cdot \delta(\tau - \tau_p),
\label{eq:ddsf}
\vspace{-1ex}
\end{equation}
where the time- and frequency-selectivity characteristics are observed in the form of unique impulses in the delay-Doppler plane corresponding to each propagation path, as illustrated in Fig. \ref{fig:DVIRF}.

\newpage

It is important to note that the equivalent channel models in eqs. \eqref{eq:TVIR} - \eqref{eq:ddsf} are based on a practical approximation of the true doubly-dispersive wireless channel, represented in terms of a finite number of significant delay and Doppler frequency taps\footnote{Strictly speaking, a doubly-dispersive channel cannot be represented by both finite delay and Doppler taps, due to the incompatibility of the bandwidth-limited assumption and the temporally-limited assumption of the signals.
Still, the approximation is well adopted in relevant research and studies \cite{Bemani_TWC23,Hong_2022}.}, which is known to generally work well \cite{Hong_2022, Bliss_13} epsecially under the underspread environment assumption, \textit{i.e.,} the maximum delay spread $\tau_\mathrm{max} - \tau_\mathrm{min}$ is smaller than $T$, the maximum Doppler spread $\nu_\mathrm{max} - \nu_\mathrm{min}$ is smaller than $\frac{1}{T}$, and $\tau_\mathrm{max} \nu_\mathrm{max} <\!< 1$, where $T$ is the finite signal period in seconds.

\begin{figure}[t]
\centering
\includegraphics[width=0.95\textwidth]{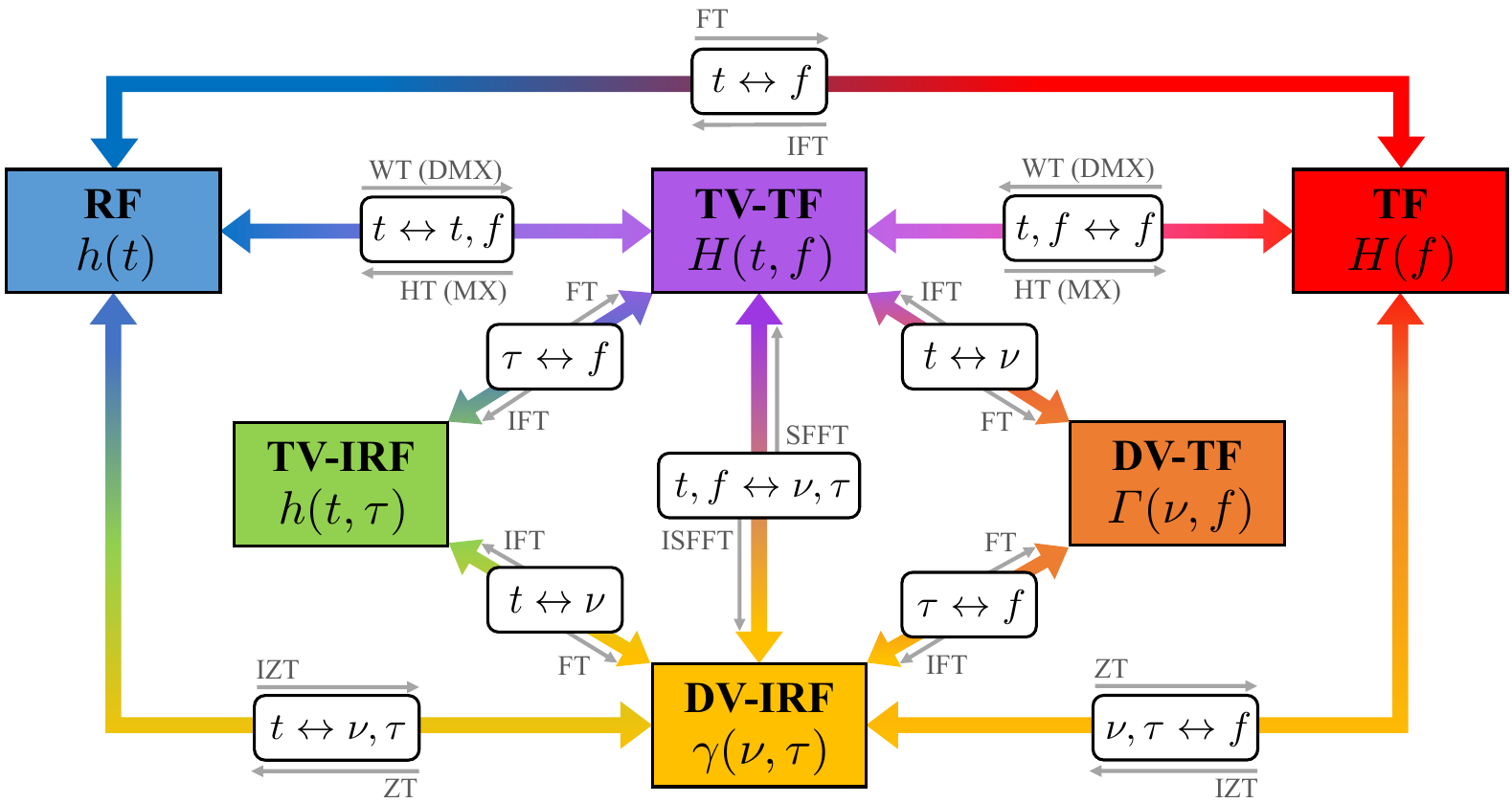}
\caption{Illustration of the relationship between the different signal domains and inherent transforms.
In addition to the illustrated domains, a special case of the time-frequency domain, namely the \textbf{chirp domain}, also exists, which is omitted in this diagram but elaborated upon in the following sections.}
\label{fig:domain_transform_map}
\vspace{-3ex}
\end{figure}

Following the above, Fig. \ref{fig:domain_transform_map} provides a diagram that summarizes the relationships between the various signal domains, including the direction and the integral domain of the necessary linear transforms.
Specifically, the rhombus-shaped relationship at the center of the figure illustrates the different domain representations of the doubly-dispersive channel as described above, which also includes the omitted \ac{DVTF}\footnote{The Doppler-frequency domain \ac{DVTF} is not addressed as often compared to the other three forms, due to its lesser intuitive relationship with the physical phenomena. However, it is still an equally valid representation of the doubly-dispersive channel.} in the Doppler-frequency domain.

The \ac{WT} and the \ac{HT} (illustrated in red) are generalizations of the \ac{MX} and \ac{DMX} operations of the classical \ac{OFDM} modulator, which transform a \ac{2D} time-frequency domain signal into the single time and frequency domains.
As will be discussed in the following section, the two transforms can be respectively implemented using the \ac{DFT} and the \ac{IDFT}.
Furthermore, as can be seen in the figure, there also exist linear transforms that directly describe concatenated \acp{FT} and/or \acp{IFT}.
Such linear transforms, such as the \ac{SFFT} and \ac{ZT}, are leveraged in the transmitter design of next-generation waveforms such as the \ac{OTFS} \cite{Wei_WC21}, and are elaborated in Sec. \ref{sec:waveforms}.

To wrap up the signal domain fundamentals, let us also address the \acf{LCT} \cite{Healy_LCT15}, also known as the \ac{AFT}, which is a four-parameter transform generalizing\footnote{Setting specific parameters reduces the \ac{LCT} to the classical transforms such as $(0, \frac{1}{2\pi}, -2\pi, 0)$ for the \ac{FT}, $(0, \frac{j}{2\pi}, j2\pi,0)$ for the \ac{LT}, and $(\mathrm{cos}\theta, \frac{1}{2\pi}\mathrm{sin}\theta, -2\pi\mathrm{sin}\theta,\mathrm{cos}\theta)$ to yield the $\theta$-th order \ac{frFT}.} many of the popular transforms such as the \ac{FT}, \acf{LT}, and \acf{fnT}.

In particular, the \ac{AFT} $\!\underset{t \rightarrow u}{\mathcal{L}\!}[\;\cdot\;]$ of a time-domain signal $s(t)$ is described by
\begin{equation}
\underset{t \rightarrow u}{\mathcal{L}}\big[s(t)\big] \triangleq
\begin{cases}
\displaystyle\int^{+\infty}_{-\infty} \! s(t) \cdot \frac{1}{\sqrt{2\pi|b|}}\cdot e^{-j(\frac{a}{2b}u^2 + \frac{1}{b}ut + \frac{d}{2b}t^2)} \;{\rm d}t,  & b \neq 0, \\
s(d \cdot u) \cdot \dfrac{1}{\sqrt{a}} \cdot e^{-j\frac{cd}{2}u^2} & b = 0, \\
\end{cases}
\label{eq:LCT}
\end{equation}
where the four \ac{AFT} parameters $(a,b,c,d)$ are arbitrary complex scalars satisfying $ad - bc = 1$.

As shall be seen, the available \ac{DoF}s of the \ac{AFT} are exploited by another promising waveform, \ac{AFDM} \cite{Bemani_TWC23}, which allows for the optimization of the \ac{AFT} parameters based on the channel statistics in order to ensure the orthogonality of the signal in doubly-dispersive systems.

\subsection{Input-Output Relationship of Doubly-Dispersive Channel}
\label{subsec:IOR_DDchannel}

The received signal model over the doubly-dispersive wireless channel $h(t, \tau)$ in response to a time-domain transmit signal $s(t)$ with bandwidth $B$, is classically described by the linear convolution over the delay domain leveraging the \ac{TVIRF} representation \cite{Hong_2022, Bliss_13}, namely
\begin{equation}
r(t) = s(t) \ast h(t, \tau) + w(t) \triangleq \int_{-\infty}^{+\infty} s(t - \tau)\bigg(\sum_{p=1}^{P} h_p \cdot e^{j2\pi \nu_p t} \cdot \delta(\tau - \tau_p) \!\bigg) \!\;{\rm d}\tau + w(t),
\label{eq:IO_continuous}
\end{equation}
where and $r(t)$ and $w(t)$ are respectively the received signal and \ac{AWGN} in time, while $\ast$ denotes the linear convolution operator.

In turn, assuming that the channel and the \ac{AWGN} are also bandlimited by bandwidth $B$ of the transmit signal, sampling eq. \eqref{eq:IO_continuous} at a sampling rate of $f_\mathrm{s} \triangleq \frac{1}{T_\mathrm{s}}$ [Hz] yields the discretized equivalent signal
\begin{equation}
r(nT_\mathrm{s}) = {\sum_{\ell = 0}^{\infty}}  s(nT_\mathrm{s}- \ell T_\mathrm{s}) \bigg( \sum_{p = 1}^{P} h_p \cdot e^{j2\pi {\nu_p} nT_\mathrm{s}} \cdot \mathrm{sinc}\big(\ell - \tfrac{\tau_p}{T_\mathrm{s}}\big) \bigg) + w[n],
\label{eq:IO_sampled}
\end{equation}
where $n \in \{0,\cdots\!,N\!-\!1\}$ and $\ell \in \{0,\cdots\!,N\!-\!1\}$ are the discrete time and delay indices, $T_\mathrm{s}$ [s] is the sampling period, and $\mathrm{sinc}(x) \triangleq \frac{\mathrm{sin}(\pi x)}{\pi x}$ is the normalized sinc function. 

The prescence of sinc functions resulting from sampling with a finite spectral support can result in interference between the Doppler responses between unique delay taps.
However, assuming a wideband communication system as expected in the considered \ac{B5G} and \ac{6G} scenarios, and in aid of oversampling if necessary, the sampling rate $f_\mathrm{s}$ will be sufficiently high such that the normalized path delays $\ell_p \triangleq \tau_p f_\mathrm{s} = \frac{\tau_p}{T_\mathrm{s}}$ can be rounded to the nearest integer with negligible error, \textit{i.e.,} $\ell_p - \lfloor\frac{\tau_p}{T_\mathrm{s}}\rceil \approx 0$, or in other words, the sampling resolution $T_\mathrm{s}$ is assumed to be sufficiently high such that the normalized path delays $\ell_p$ are integers \cite{Hong_2022,Bemani_TWC23,Wei_WC21}.
Consequently, the sinc functions are then actually equivalent to unit impulse functions\footnote{Given that the normalized path delays $\ell_p$ are only integer-valued, the sinc functions which take integer input $(\ell - \ell_p)\in\mathbb{N}_0$, yield a value of 1 iff $\ell - \ell_p = 0$, and a value of $0$ otherwise $-$ this is equivalent to the discrete unit impulse function $\delta[\ell - \ell_p]$.}, such that eq. \eqref{eq:IO_sampled} can be expressed in terms of the sampled sequences
\begin{equation}
r[n] = {\sum_{\ell = 0}^{\infty}}  s[n - \ell] \bigg( \sum_{p = 1}^{P} h_p \cdot e^{j2\pi f_p \frac{n}{N}} \cdot {\delta \big[ \ell - \ell_p \big]} \bigg) + w[n],
\label{eq:IO_discrete}
\end{equation}
where $r[n]$, $s[n]$, and $w[n]$ are respectively the sampled sequences of $r(t)$, $s(t)$, and $w(t)$; $f_p \triangleq \frac{N \nu_p}{f_\mathrm{s}} \in [-\frac{N\nu_\mathrm{max}}{f_\mathrm{s}}, \frac{N\nu_\mathrm{max}}{f_\mathrm{s}}]$ is the normalized digital Doppler shift of the $p$-th path; $\ell_p \triangleq \frac{\tau_p}{T_\mathrm{s}} \in \{0,\cdots\!,\ell_{\mathrm{max}}\}$ is the normalized integer delay of the $p$-th path; and $\delta[\,\cdot\,]$ is the discrete unit impulse function.

In addition, in practical multicarrier wireless communications techniques, the transmit sequence in eq. \eqref{eq:IO_discrete} is prepended with a \acf{CP} to mitigate the effects of time dispersio
The prefix sequence is defined within a \ac{CP} length of $N_\mathrm{cp}$ samples, with $N_\mathrm{cp} \geq \ell_\mathrm{max}$ such that
\begin{equation} 
s[n'] = s[N + n'] \cdot e^{j2\pi \cdot \phi_\mathrm{cp}(n')},
\label{eq:cyclic_prefix}
\end{equation}
where $n' \in \{-1,\cdots,-N_\mathrm{cp}\}$, and $\phi_\mathrm{cp}(n')$ is a function denoting the multiplicative phase term specific for each waveform, which for example, is set to zero if the \ac{CP} does not require a phase offset, as in the \ac{OFDM}, or a chirp-based phase offset as in the \ac{AFDM} as will be seen in the following sections.

The \ac{CP} as described in eq. \eqref{eq:cyclic_prefix} enables the linear convolutional input-output relation of the \ac{TVIRF} to be processed as a circular convolutional response.
After removing the received signal parts corresponding to the \ac{CP}, the circular convolutional input-output relationship can be described in matrix form as
\begin{equation}
\mathbf{r} \triangleq \mathbf{H} \!\cdot\! \mathbf{s} = \Big( \sum_{p=1}^{P} {\overbrace{h_p \!\cdot\! \mathbf{\Phi}_{p} \!\cdot\! \mathbf{W}^{f_p} \!\cdot\! \mathbf{\Pi}^{\ell_p}}^{\triangleq \mathbf{H}_p \in \mathbb{C}^{N \times N}}}\,\Big)\!\cdot \;\! \mathbf{s} + \mathbf{w} = {\Big(\!\sum_{p=1}^{P} \mathbf{H}_p\Big) \!\cdot\mathbf{s} + \mathbf{w}} \in \mathbb{C}^{N \times 1},
\label{eq:IO_matrix}
\vspace{-1ex}
\end{equation}
where $\mathbf{r} \!\in\! \mathbb{C}^{N \times 1}$, $\mathbf{s} \!\in\! \mathbb{C}^{N \times 1}$, and $\mathbf{w} \!\in\! \mathbb{C}^{N \times 1}$ are respectively the vectors representing the received signal, the transmit signal, and \ac{AWGN}; 
$\mathbf{H} \in \mathbb{C}^{N \times N}$ is the circular convolution effective channel matrix; 
$\mathbf{\Phi}_{p} \!\in\! \mathbb{C}^{N \times N}$ is the diagonal matrix corresponding to the $p$-th delayed \ac{CP} phase as given in eq. \eqref{eq:CP_phase_matrix};
$\mathbf{W} \!\in\! \mathbb{C}^{N \times N}$ is the diagonal matrix containing the $N$-th roots of unity as given in eq. \eqref{eq:W_matrix} whose exponentiation by a real number is equivalent to the element-wise exponentiation of the diagonal elements;
and $\mathbf{\Pi} \in \mathbb{C}^{N \times N}$ is the forward cyclic shift matrix\footnote{To clarify with an example of $N = 3$, $\mathbf{\Pi}^0 \equiv \mathbf{I}_3 = {\scalefont{0.4}
\begin{bmatrix} 
1~~0~~0\\[-1ex]
0~~1~~0\\[-1ex]
0~~0~~1\\
\end{bmatrix}}$, 
$\mathbf{\Pi}^1 = {\scalefont{0.4}
\begin{bmatrix} 
0~~0~~1\\[-1ex]
1~~0~~0\\[-1ex]
0~~1~~0\\
\end{bmatrix}}$, and
$\mathbf{\Pi}^2 = {\scalefont{0.4}
\begin{bmatrix} 0~~1~~0\\[-1ex]
0~~0~~1\\[-1ex]
1~~0~~0\\
\end{bmatrix}}$, and so on.
}
obtained by left-shifting the $N \times N$ identity matrix once, such that right-multiplying by $\mathbf{\Pi}^{\ell_p}$ corresponds to a cyclic left-shift of a matrix by $\ell_p \in \mathbb{N}_0$ elements. 

\begin{figure*}[b]
\vspace{-4ex}
\hrule
\vspace{0.5ex}
\begin{equation}
\mathbf{\Phi}_{p} \triangleq \mathrm{diag}\Big(\big[\overbrace{e^{-j2\pi\cdot \phi_\mathrm{cp}(\ell_p)}, e^{-j2\pi\cdot \phi_\mathrm{cp}(\ell_p - 1)}, \cdots, e^{-j2\pi\cdot \phi_\mathrm{cp}(2)}, e^{-j2\pi\cdot \phi_\mathrm{cp}(1)}}^{\ell_p \;\text{terms}}, \overbrace{\;\!1\;\!, 1\;\!, \cdots\!\vphantom{e^{(x)}}, 1\;\!, 1}^{N - \ell_p\;\! \text{ones}}\big]\Big) \in \mathbb{C}^{N \times N}\!.
\label{eq:CP_phase_matrix}
\vspace{-1ex}
\end{equation}
\begin{equation}
\mathbf{W} \triangleq \mathrm{diag}\Big(\big[1, e^{-j2\pi/N}, \cdots, e^{-j2\pi(N-2)/N}, e^{-j2\pi(N-1)/N}\big]\Big) \in \mathbb{C}^{N \times N}.
\label{eq:W_matrix}
\end{equation}
\vspace{-4ex}
\end{figure*}

As can be seen in eq. \eqref{eq:IO_matrix}, the input-output relationship of a doubly-dispersive channel is described by a matrix $\mathbf{H}$ consisting of $P$ off-diagonals, whose shifted positions are determined by the integer delay of each path.
In addition, the complex values along each diagonal contain the channel fading coefficient and the phase offset information of the delayed \ac{CP} and the Doppler shift of each path.
As a consequence of the circulant convolutional channel structure, the different paths are only resolvable in the delay domain and not in the Doppler domain.
Therefore, a key design objective of a double-dispersion robust waveform must be the ability to orthogonalize the delays and Doppler shifts of the channel via means of novel domain transforms in the modulation and demodulation of the transmit signal.

\subsection{\acs{MIMO} System Model of Doubly-Dispersive Channels}
\label{sec:MIMO_channel model}
Extending the \ac{SISO} model of the doubly-dispersive channel and its received signal model to the \ac{MIMO} scenario between a transmitter array equipped $N_T$ antennas and a receiver array equipped with $N_R$ antennas, the received signal $\mathbf{r}_{n_r} \in \mathbb{C}^{N \times 1}$ at the $n_r$-th receive antenna, with $n_r \in \{1,\cdots,N_R\}$, is given by
\begin{equation}
\mathbf{r}_{n_r} \triangleq \sum_{n_t = 1}^{N_T} \mathbf{H}_{n_t,n_r} \!\!\cdot\! \mathbf{s}_{n_t} + \mathbf{w}_{n_r} = \sum_{n_t = 1}^{N_T} \!\Big(\! \overbrace{\sum_{p=1}^{P} h_{p,n_t,n_r} \!\cdot\! \mathbf{\Phi}_{p} \!\cdot\! \mathbf{W}^{f_p} \!\cdot\! \mathbf{\Pi}^{\ell_p}}^{\triangleq \mathbf{H}_{n_t, n_r}}\Big) \!\cdot\! \mathbf{s}_{n_t} + \mathbf{w}_{n_r} \in \mathbb{C}^{N \times 1},
\label{eq:received_signal_MIMO} 
\end{equation}
where $\mathbf{H}_{n_t, n_r} \in \mathbb{C}^{N \times N}$ is the convolutional channel between the $n_t$-th transmit antenna and the $n_r$-th receive antenna, with $n_t \in \{1,\cdots,N_T\}$, $\mathbf{s}_{n_t} \in \mathbb{C}^{N \times 1}$ is the transmit signal vector from the $n_t$-th transmit antenna, $h_{p,n_t,n_r} \in \mathbb{C}$ is the complex channel fading coefficient corresponding to the propagation path between the $n_t$-th transmit antenna and the $n_r$-th receive antenna via the $p$-th scatter, and $\mathbf{w}_{n_r} \in \mathbb{C}^{N \times 1}$ is the \ac{AWGN} vector at the $n_r$-th receive antenna.

Notice that the \ac{MIMO} received signal model in eq. \eqref{eq:received_signal_MIMO} inherently assumes that the path delays $\ell_p$ and Doppler shifts $f_p$ are not spatially dependent, \textit{i.e.,} they are identical across all transmit and receive antenna pairs for each propagation path, in accordance to the \ac{SotA} standardization reports and literature \cite{ETSI_125996}.
Therefore, only the complex fading coefficients $h_{p,n_t,n_r} \in \mathbb{C}$ are considered to be dependent on the specific $p$-th path between the $n_t$-th transmit antenna and the $n_r$-th receive antenna.

However, when beamforming is considered at the \ac{MIMO} arrays, the effects of spatial filtering \cite{Li_Book08} introduce angle-dependent antenna gains at the transmitter and/or receiver, which may vary the visibility and hence the  doubly-dispersive channel statistics.
To this end, consider an arbitrary transmit beamformer and a receive beamformer applied at the \ac{MIMO} transmitter and receiver, respectively, where the corresponding transmit beamforming gain at an angle-of-departure $\theta^{\mathrm{t}}$ be denoted by $f(\theta^{\mathrm{t}})$, and the receive beamforming gain at an angle-of-arrival $\theta^{\mathrm{r}}$ by $g(\theta^{\mathrm{r}})$.
Then, the beamformed channel between the $n_t$-th transmit antenna and the $n_r$-th receive antenna is described by
\begin{equation}
\mathbf{H}_{n_t,n_r} = \sum_{p=1}^{P} g(\theta^{\mathrm{r}}_p) \!\cdot\! f(\theta^{\mathrm{t}}_p) \!\cdot\! h_{p,n_t,n_r} \!\cdot\!\big (\mathbf{\Phi}_{p} \!\cdot\! \mathbf{W}^{f_p} \!\cdot\! \mathbf{\Pi}^{\ell_p} \! \big),
\label{eq:beamform_channel}
\end{equation}
where $\theta^{\mathrm{t}}_p$ and $\theta^{\mathrm{r}}_p$ are the relative angle of the $p$-th scatterer to the transmitter and the receiver arrays\footnote{This model can be generalized to \ac{2D} planar arrays incorporating the azimuth and elevation angles of arrival and departure.}. 

Due to the ``direction''-dependent gains of the transmit and receive beamformers for each path, it is implied that a given scatterer and its doubly-dispersive propagation path is only visible to the channel if it is within the angular range of the major beam.
Consequently, the prescence of certain scattering paths may be rendered negligible, such that for \ac{MIMO} systems, the doubly-dispersive channel statistics including delay spread, Doppler spread, and the number of significant paths, are beamspace-dependent.

\section{Signal Models of Next-Generation Waveforms}
\label{sec:waveforms}

In this section, we provide the signal models and the transmitter structures of the various waveforms proposed for high performance in the doubly-dispersive channel.
First, the \ac{OFDM} waveform is described as a reference, followed by the next-generation waveform candidates, namely, \ac{OTFS}, and \ac{AFDM}, in addition to various derivative waveforms that can be related to the latter\footnote{A Matlab$^\text{\copyright}$ implementation of the doubly-dispersive channel model described in this section, as well as a convenient channel visualization tool used to generate some of the figures can be found our online respository \href{https://pages.constructor.university/abreugroup/doubly-dispersive-channel-simulation-package/}{[\textit{here}]}.}.

\vspace{-1ex}
\subsection{Orthogonal Frequency Division Multiplexing (\ac{OFDM})}
\label{subsec:OFDM}

The well-known \ac{OFDM} transmitter modulates digital symbols from the frequency domain into a time domain signal employing an \ac{IDFT} operation, as illustrated in Fig. \ref{fig:domain_transform_map}.
Namely, given a vector $\mathbf{x} \in \mathbb{C}^{N \times 1}$ consisting of $N$ complex symbols, the \ac{OFDM} transmit signal $\mathbf{s}^\mathrm{OFDM}\in \mathbb{C}^{N \times 1}$ is given by \cite{Prasad_04}
\begin{equation}
\mathbf{s}^\mathrm{OFDM} = \mathbf{F}^{-1}_N \cdot \mathbf{x} \in \mathbb{C}^{N \times 1},
\end{equation}
where $\mathbf{F}_N \in \mathbb{C}^{N \times N}$ is the $N$-point \ac{DFT} matrix, and hence $\mathbf{F}_N^{-1} \triangleq \mathbf{F}_N\herm$ is the $N$-point \ac{IDFT} matrix. 

Following the above, the received signal $\mathbf{y}^\mathrm{OFDM} \in \mathbb{C}^{N \times 1}$ over the circular convolutional channel described by eq. \eqref{eq:IO_matrix} is demodulated via the forward $N$-point \ac{DFT}, \textit{i.e.,} 
\begin{equation}
\mathbf{y}^\mathrm{OFDM} = \mathbf{F}_N 
\overbrace{\big( \mathbf{H} \cdot \mathbf{s}^\mathrm{OFDM}  + \mathbf{w} \big)}^{\triangleq \;\! \mathbf{r}^\mathrm{OFDM} \; \in \; \mathbb{C}^{N \times 1}} = \overbrace{(\mathbf{F}_N \cdot \mathbf{H} \cdot\mathbf{F}^{\mathsf{H}}_N )}^{\triangleq \;\! \mathbf{G}^\mathrm{OFDM} \;\! \in \;\! \mathbb{C}^{N \times N}}\mathbf{x} + \mathbf{F}_N \cdot \mathbf{w} \in \mathbb{C}^{N \times 1},
\label{eq:OFDM_txsig}
\end{equation}
where, for exposition convenience, we defined the \ac{OFDM} receive signal as $\mathbf{r}^\mathrm{OFDM} \in \mathbb{C}^{N \times 1}$, and the effective channel $\mathbf{G}^\mathrm{OFDM} \in \mathbb{C}^{N \times N}$ describing the input-output relationship of the baseband \ac{OFDM} symbols, which can be obtained by combining eqs. \eqref{eq:IO_matrix} and \eqref{eq:OFDM_txsig} to yield
\begin{equation}
\mathbf{G}^\mathrm{OFDM} \triangleq \sum_{p=1}^{P}  h_p \cdot \mathbf{F}_N  \big (\mathbf{W}^{f_p} \!\cdot\! \mathbf{\Pi}^{\ell_p} \! \big) \mathbf{F}\herm_N,
\label{eq:eff_ofdm}
\end{equation}
where the \ac{CP} phase matrices $\mathbf{\Phi}_p$ existent in eq. \eqref{eq:IO_matrix} have been reduced to identity matrices, as the \ac{CP} of \ac{OFDM} signals does not require a phase offset, \textit{i.e.,} $\phi_\mathrm{cp}(n) = 0$ in eq. \eqref{eq:CP_phase_matrix}.

As illustrated in Fig. \ref{fig:effChannels}, the column-wise \ac{DFT} and row-wise \ac{IDFT} in presence of fractional normalized digital Doppler shifts cause the channel diagonals of the convolution matrix to be spread into a decaying band, centered at the original diagonals, such that significant interference between each path may arise.

\subsection{Orthogonal Time Frequency Space (\ac{OTFS})}
\label{subsec:OTFS}

In the \ac{OTFS} modulation scheme, the information symbols are first directly placed in the delay-Doppler domain, instead of the frequency domain as in the \ac{OFDM} approach, which are then multiplexed into the time signal.
Namely, the complex baseband symbols are structured into a \ac{2D} grid of size $K \times L$ in the delay-Doppler domain, which is transformed into a time-domain signal via the \ac{IDZT}.
Alternatively, as illustrated in Fig. \ref{fig:domain_transform_map}, the equivalent\footnote{While mathematically equivalent, in practice, the two-step \ac{SFFT}-based transform process has been shown to exhibit higher Doppler leakage than the \ac{ZT}-based counterpart, whose effect decreases with a larger number of subcarriers \cite{Bhat_CL23}.} domain transform can be achieved via a two-step process, in which the delay-Doppler signal is first transformed into the time-frequency domain via the \ac{ISFFT}, and then into the continuous time signal via a pulse-shaping \ac{HT}.

This article will hereafter adopt the more common \ac{ISFFT} formulation implemented via \acp{DFT}/\acp{IDFT}.
In light of the above, the \ac{OTFS} modulation process can be described mathematically as
\begin{align}
\mathbf{s}^{\mathrm{OTFS}} \triangleq  \mathrm{vec} \Big( \hspace{-2ex} \overbrace{\mathbf{P}^{\mathrm{tx}} \mathbf{F}^{-1}_{\!K}}^{\text{Pulse-shaping HT}} \hspace{-2ex} \cdot \overbrace{ \big( \mathbf{F}^{}_{\!K} \mathbf{X} \mathbf{F}_{\!L}^{-1} \big) }^{\text{ISFFT}} \Big) 
= (\mathbf{F}^{-1}_{\!L} \otimes \mathbf{P}^{\mathrm{tx}}) \cdot \overbrace{\mathrm{vec}(\mathbf{X})}^{\triangleq \mathbf{x}} \in \mathbb{C}^{KL \times 1}, 
\label{eq:tx_otfs}
\end{align} 
\vspace{-5.5ex}

\noindent where $\mathbf{s}^\mathrm{OTFS} \in \mathbb{C}^{KL \times 1}$ is the \ac{OTFS} transmit signal vector, $\mathbf{X} \in \mathbb{C}^{K \times L}$ is the information symbol matrix consisting of $N \triangleq KL$ number of complex symbols\footnote{Without loss of generality, we assume $N \triangleq KL$ to enable direct comparison with \acs{1D} modulation schemes, \textit{i.e.,} \ac{OFDM}.}, $\mathbf{P}^{\mathrm{tx}} \in \mathbf{C}^{K \times K}$ is the diagonal transmit pulse-shaping filter matrix\footnote{The literature commonly assumes a non-ideal rectangular pulse-shape, which reduces $\mathbf{P}^\mathrm{tx}$ to a $K \times K$ identity matrix.
However, the specific design of the \ac{OTFS} pulse-shaping function, in addition to the \ac{OTFS} time-frequency windowing function \cite{Wei_TC21}, is an important design criteria for \ac{OTFS} which must be addressed to improve the communciations and sensing performance.}, $\mathbf{F}_{\!K} \in \mathbb{C}^{K \times K}$ and $\mathbf{F}_{\!L} \in \mathbb{C}^{L \times L}$ are the $K$-point and $L$-point \ac{DFT} matrices, and $\mathrm{vec}(\cdot)$ and $\otimes$ denote the stacking vectorization and Kronecker product operators, respectively.

The filtered and demodulated signal $\mathbf{y}^{\mathrm{OTFS}}$ after the convolution channel $\mathbf{H}$ in eq. \eqref{eq:IO_matrix} is given by \vspace{-1ex}
\begin{equation}
\mathbf{y}^{\mathrm{OTFS}} \triangleq (\mathbf{F}_L \otimes \mathbf{P}^{\mathrm{rx}}) \overbrace{\big(\mathbf{H} \!\cdot \!\mathbf{s}^{\mathrm{OTFS}} \!+ \mathbf{w}\big)}^{\triangleq \; \mathbf{r}^\mathrm{OTFS} \; \in \; \mathbb{C}^{KL \times 1}  } = \mathbf{G}^{\mathrm{OTFS}} \!\cdot\! \mathbf{x} + (\mathbf{F}_L \otimes \mathbf{P}^{\mathrm{rx}}) \!\cdot\! \mathbf{w} \in \mathbb{C}^{KL \times 1},
\label{eq:rx_otfs}
\vspace{-2ex}
\end{equation}
where $\mathbf{P}^{\mathrm{rx}} \in \mathbb{C}^{K \times K}$ is the diagonal matched filter matrix of $\mathbf{P}^\mathrm{tx}$, and the effective \ac{OTFS} channel $\mathbf{G}^{\mathrm{OTFS}} \in \mathbb{C}^{N \times N}$ in the delay-Doppler domain is given by \vspace{-1ex}
\begin{equation}
\mathbf{G}^{\mathrm{OTFS}} \triangleq \sum_{p=1}^{P} h_p \cdot \big(\mathbf{F}_L^{~} \otimes \mathbf{P}^{\mathrm{rx}}\big) \Big(\mathbf{W}^{f_p} \!\cdot\! \mathbf{\Pi}^{\ell_p} \Big) \big(\mathbf{F}\herm_{L} \otimes \mathbf{P}^{\mathrm{tx}}\big) \in \mathbb{C}^{N \times N},
\label{eq:eff_otfs}
\vspace{-1ex}
\end{equation}
which is the convolutional channel matrix $\mathbf{H}$ after a block-wise pulse-shaped \acs{DFT} and \acs{IDFT}, and similarly to the \ac{OFDM} waveform, the \ac{CP} phase matrices $\mathbf{\Phi}_p$ have been reduced to identity matrices, as the \ac{OTFS} signal also does not require a phase offset.

It can be observed from eq. \eqref{eq:eff_otfs} that the $KL = N$ elements in each diagonal of the convolutional matrix of eq. \eqref{eq:IO_matrix} are spread into the \ac{OTFS} effective channel via the block-wise pulse-shaping \acp{FT}, such that the $KL \times KL$ \ac{OTFS} channel matrix can be considered as a $K \times K$ grid of $L \times L$ sub-matrices (illustrated as minor grids in Fig. \ref{fig:effChannels}).
In light of the above, the positions of the non-zero channel coefficient elements can be deterministically obtained by the values of both normalized delay and normalized digital Doppler of each path, respective to the occupied sub-matrices and the amount of left-shift of the diagonals.

If the normalized digital Doppler frequencies are integers, \textit{i.e.,} $f_p \in \mathbb{Z}$, (\ul{not} the true Doppler frequencies $\nu_p$), each path occupies exactly $K$ sub-matrices out of the $K \times K$ grid in a shifted block-diagonal structure, whose amount of right-shift is determined by the integer component of the Doppler shift, $f_p^\mathrm{int} \triangleq \lfloor f_p \rceil$.
For example, a path with $f_p^\mathrm{int} = 0$ occupies the $K$ sub-matrices in the main block-diagonal (shift of index $0$), whereas a path with $f_p^\mathrm{int} = 1$ occupies the $K$ sub-matrices in the block-diagonal, which is right-shifted by an index of 1.
On the other hand for negative Doppler shifts, the block-diagonals are left-shifted by an index of $|f_p^\mathrm{int}|$, as illustrated by path $3$ in Fig. \ref{fig:effChan_Int}.

In turn, each of the $K$ occupied sub-matrices follows the same structure consisting of exactly $L$ non-zero elements in a shifted diagonal, with a left-shift relative to the main diagonal determined by the value of the normalized path delay $\ell_p \in \mathbb{N}$.
For example, for a path with delay $\ell_p = 0$, the $L$ non-zero elements are placed in the main diagonal for all $K$ sub-matrices, whereas a path with $\ell_p = 3$ will have the $L$ non-zero elements in the diagonal left-shifted by three indices, for all $K$ sub-matrices.

It follows that the \ac{OTFS} waveform can achieve complete orthogonality and resolvability in the delay-Doppler domain assuming integer values of the normalized delay and normalized digital Doppler shifts, and if the channel satisfies the orthogonality condition given by $\ell^\mathrm{max} \leq L - 1$ and $f^\mathrm{max} \leq \lfloor\frac{K}{2}\rfloor$, where $\ell^\mathrm{max} \triangleq \big\lceil\frac{\tau^\mathrm{max}}{T_\mathrm{s}}\big\rceil$ and $f^\mathrm{max} \triangleq \big\lfloor \frac{N\nu^\mathrm{\max}}{f_\mathrm{s}} \big\rfloor$ are the maximum normalized delay and digital Doppler shift.
Note that the orthogonality condition inherently implies that the \ac{OTFS} waveform will remain orthogonal between unique paths even when the Doppler spread is negligible, \textit{i.e.,} $f^\mathrm{max} = 0$, for fixed $K$ and $L$.

In contrast, in the case of \textbf{fractional} values of normalized digital Doppler, \textit{i.e.,} $f_p \triangleq f_p^\mathrm{int} + f_p^\mathrm{frac} \in \mathbb{R}$, where $f_p^\mathrm{frac} \in [0.5, +0.5)$, the powers of the channel elements are diffused (or ``leaked'') across all $K^2$ sub-matrices over the main $K$ block-diagonal sub-matrices of the integer case, resulting in a Doppler domain interference as illustrated in Fig. \ref{fig:effChan_Frac} and Fig. \ref{fig:effChan_Frac3D}.
The amount of such power leakage is determined by the magnitude of the fractional component of the normalized digital Doppler shift $|f^\mathrm{frac}_p|$, such that larger magnitudes result in more leakage nd hence a more severe interference.

\vspace{-2ex}
\subsection{Affine Frequency Division Multiplexing (\ac{AFDM})} 
\label{subsec:AFDM}

In \ac{AFDM}, a one-dimensional vector of symbols $\mathbf{x} \in \mathbb{C}^{N \times 1}$ is directly multiplexed into a \textit{twisted} time-frequency chirp domain using the \ac{IDAFT} \cite{Pei_TSP00}, as described by \vspace{-1ex}
\begin{equation}
\mathbf{s}^{\mathrm{AFDM}} \triangleq \mathbf{A}^{\!-1} \!\cdot\! \mathbf{x} = \overbrace{(\mathbf{\Lambda}_{c_2} \!\!\cdot\! \mathbf{F}_N \!\cdot\! \mathbf{\Lambda}_{c_1} )^{-1}}^{\text{IDAFT}} \!\cdot \;\! \mathbf{x} =  (\mathbf{\Lambda}_{c_1}\herm \!\cdot\! \mathbf{F}_N\herm \!\cdot\! \mathbf{\Lambda}_{c_2}\herm ) \!\cdot\! \mathbf{x} \in \mathbb{C}^{N \times 1},
\label{eq:tx_afdm}
\vspace{-2ex}
\end{equation} 
where $\mathbf{A} \triangleq \mathbf{\Lambda}_{c_2} \mathbf{F}_N \mathbf{\Lambda}_{c_1} \in \mathbb{C}^{N \times N}$ is the forward $N$-point \ac{DAFT} matrix, $\mathbf{\Lambda}_{c_i} \triangleq \mathrm{diag}[e^{-j2\pi c_i (0)^2}, \cdots, e^{-j2\pi c_i (N-1)^2}] \in \mathbb{C}^{N \times N}$ is a diagonal chirp matrix with a central digital frequency of $c_i$, and the central frequencies $c_1$ and $c_2$ of the two diagonal chirps\footnote{It is shown in \cite{Bemani_TWC23} that the chirp frequencies $c_1$ and $c_2$ are actually correspondent to the four configurable parameters of the \ac{AFT} formulation in eq. \eqref{eq:LCT}.} can be optimized to the channel statistics to improve the delay-Doppler orthogonality of the \ac{AFDM} effective channel.

It follows that the demodulated \ac{AFDM} signal over the convolution channel in eq. \eqref{eq:IO_matrix} is given by \vspace{-1ex}
\begin{equation}
\mathbf{y}^{\mathrm{AFDM}} = \mathbf{A} \!\cdot\! \overbrace{(\mathbf{H} \!\cdot\! \mathbf{s}^{\mathrm{AFDM}} + {\mathbf{w}})}^{\triangleq \mathbf{r}^\mathrm{AFDM} \; \in \; \mathbb{C}^{N \times 1} } = \mathbf{G}^{\mathrm{AFDM}}\!\cdot\!\mathbf{x} +  \mathbf{A} \!\cdot\! \mathbf{w} \in \mathbb{C}^{N \times 1},
\label{eq:rx_afdm}
\vspace{-2ex}
\end{equation} 
with the effective \ac{AFDM} channel given by \vspace{-1ex}
\begin{equation}
\mathbf{G}^{\mathrm{AFDM}} \triangleq \sum_{p=1}^{P}  h_p \!\cdot\! \big(\mathbf{\Lambda}_{c_2} \!\!\cdot\! \mathbf{F}_N \!\cdot\! \mathbf{\Lambda}_{c_1} \big)  \Big (\mathbf{\Phi}_{p} \!\cdot\! \mathbf{W}^{f_p} \!\cdot\! \mathbf{\Pi}^{\ell_p} \! \Big) \big(\mathbf{\Lambda}_{c_1}\herm \!\cdot\! \mathbf{F}_N\herm \!\cdot\! \mathbf{\Lambda}_{c_2}\herm \big) \in \mathbb{C}^{N \times N},
\label{eq:eff_afdm}
\vspace{-1ex}
\end{equation} 
where the \ac{CP} phase matrices $\mathbf{\Phi}_p$ require a chirp-cyclic phase offset, described by $\phi_{\mathrm{cp}}(n) = c_1(N^2 + 2Nn)$, which in the specific case of even $N$ and integer $2Nc_1$, reduces the matrices $\mathbf{\Phi}_p$ back to identity.

The \ac{AFDM} effective channel can also achieve full orthogonality in the integer normalized delay-Doppler domain when the channel satisfies the orthogonality condition, which is given by \vspace{-1ex}
\begin{equation}
2\big(f^\mathrm{max} + \xi\big)(\ell^\mathrm{max} + 1) + \ell^\mathrm{max} \leq N,
\label{eq:AFDM_orthoCondition}
\vspace{-2ex}
\end{equation}
where $f^\mathrm{max}$ and $\ell^\mathrm{max}$ are the maximum normalized digital Doppler shift and delay of the channel, $ \xi \in \mathbb{N}_0$ is a free parameter determining the so-called \textit{guard width} of the \ac{AFDM}, denoting the number of additional guard elements around the diagonals to anticipate for Doppler-domain interference.

Assuming the orthogonality condition is met, the optimal \ac{AFDM} chirp frequencies satisfy
\begin{equation}
c_1 = \frac{2(f^\mathrm{max} + \xi) + 1}{2N}, ~\;\text{and}~\;
c_2 <\!< \frac{1}{N},
\vspace{-1ex}
\end{equation}
where the flexibility in $c_2$ enables fine-tuning of the waveform shape as will be discussed in Sec. \ref{sec:radar_sensing}.

\begin{figure}[H]
\begin{subfigure}{1\textwidth}
\includegraphics[width=\linewidth]{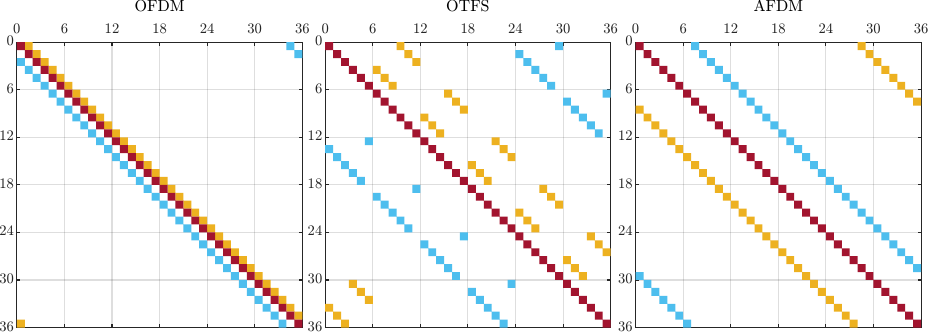}
\caption{Target parameters: \{$\ell_1 \!=\! 0$, $f_1 \!=\! 0$\} (red), \{$\ell_2 \!=\! 1$, $f_2 \!=\! -2$\} (blue), \{$\ell_3 \!=\! 3$, $f_3 \!=\! +1$\} (yellow).}
\label{fig:effChan_Int}
\end{subfigure}
\vspace{1.5ex}

\begin{subfigure}{1\textwidth}
\includegraphics[width=\linewidth]{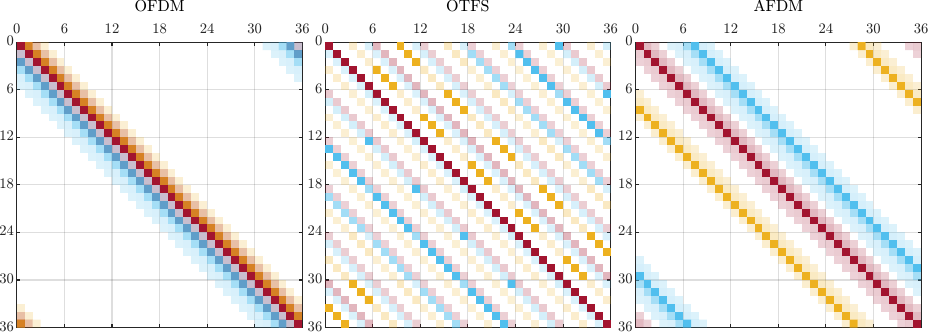}
\caption{Target parameters: \{$\ell_1 \!=\! 0$, $f_1 \!=\! 0.266$\} (red), \{$\ell_2 \!=\! 1$, $f_2 \!=\! -2.365$\} (blue), \{$\ell_3 \!=\! 3$, $f_3 \!=\! +1.231$\} (yellow).}
\label{fig:effChan_Frac}
\end{subfigure}
\vspace{1.5ex}

\begin{subfigure}{1\textwidth}
\includegraphics[width=\linewidth]{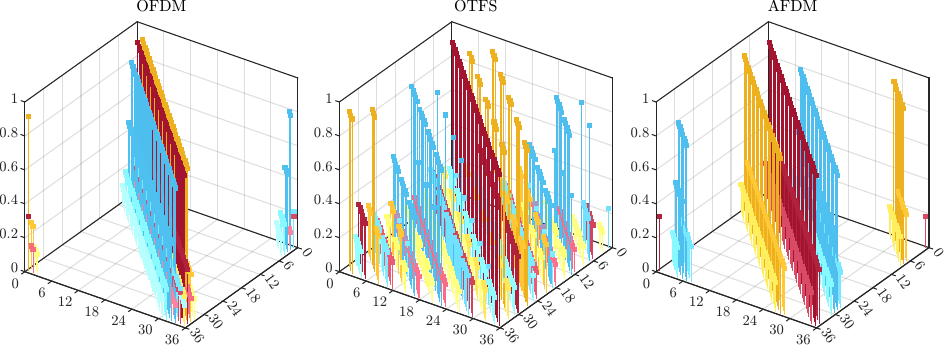}
\caption{Target parameters: \{$\ell_1 \!=\! 0$, $f_1 \!=\! 0.266$\} (red), \{$\ell_2 \!=\! 1$, $f_2 \!=\! -2.365$\} (blue), \{$\ell_3 \!=\! 3$, $f_3 \!=\! +1.231$\} (yellow).}
\label{fig:effChan_Frac3D}
\end{subfigure}
\vspace{-0.5ex}

\caption{Effective channel matrix structures of different waveforms in a doubly-dispersive channel with $P = 3$ resolvable paths (each depicted in a different color), with corresponding normalized delays $\ell_p$ and normalized digital Doppler shifts $f_p$.
The system size parameters are $N = 36$ for the \ac{OFDM} and \ac{AFDM}, and $K = 6, L = 6$ for the \ac{OTFS}.
The fading colors for the fractional Doppler case correspond to the magnitude of the elements whereas darker colors correspond to larger powers.
Channel components with a magnitude lower than $1/2N$ are considered negligible and not visualized in the figure.}
\label{fig:effChannels}
\end{figure}

In light of the above, the position of the shifted diagonal in the \ac{AFDM} channel can also be described in terms of the integer values of the normalized delay-Doppler indices of each path.
Unlike the intricate block-wise structure of the \ac{OTFS} effective channel, the \ac{AFDM} effective channel exhibits only a single diagonal per path, which is shifted by a deterministic index dependent on the integer normalized delay $\ell_p$ and integer normalized digital Doppler shift $f_p^{\mathrm{int}}$.
In other words, each diagonal of the convolution channel in eq. \eqref{eq:IO_matrix} is right-shifted by an index of exactly $\ell_p\cdot2(f^\mathrm{max} + \xi) + f_p^\mathrm{int}$ positions, as illustrated in Fig. \ref{fig:effChan_Int}.
On the other hand, in the presence of fractional components of normalized digital Doppler shifts, the diagonals of the \ac{AFDM} effective channels also exhibit a power leakage around the main diagonal, resulting in Doppler-domain interference as can be seen in Figs. \ref{fig:effChan_Frac} and \ref{fig:effChan_Frac3D}.

\subsection{Related Next-Generation Waveforms}
\label{subsec:related_waveforms}

\begin{figure}[H]
\vspace{-1.5ex}
\centering
\includegraphics[width=1\textwidth]{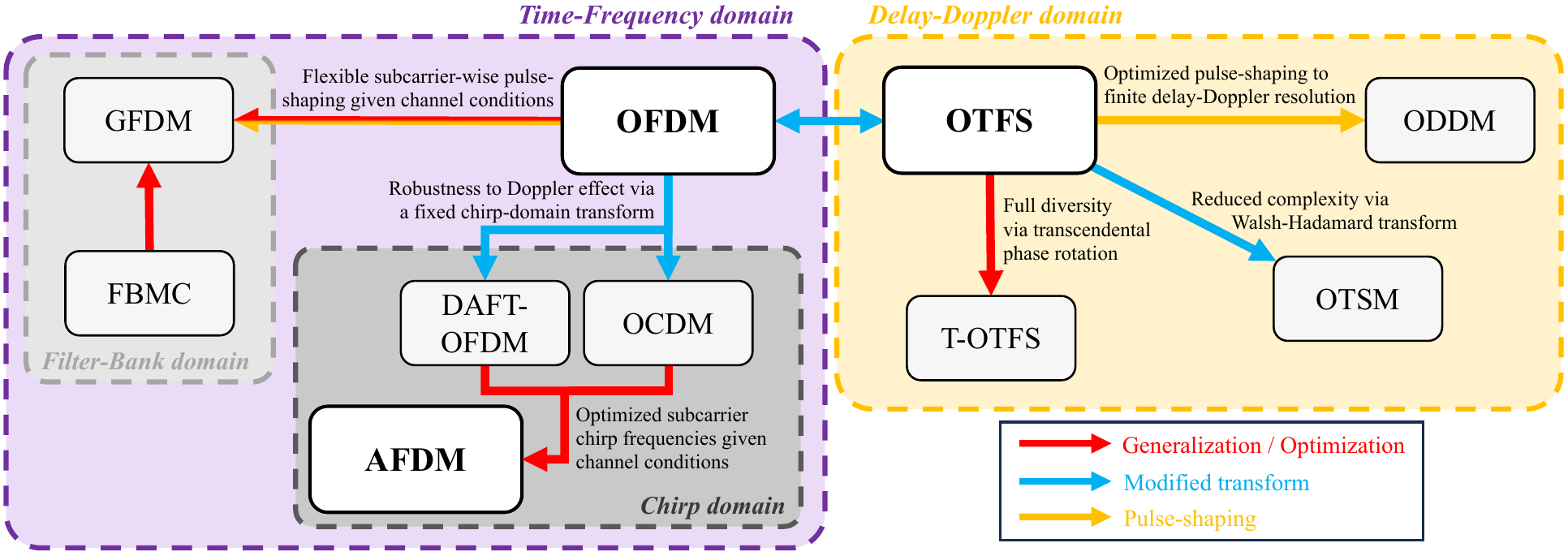}
\vspace{-3ex}
\caption{A map of relationships between next-generation waveforms and their signal domains.}
\label{fig:waveform_map}
\end{figure}

\begin{figure}[t]
\centering
\includegraphics[width=1\textwidth]{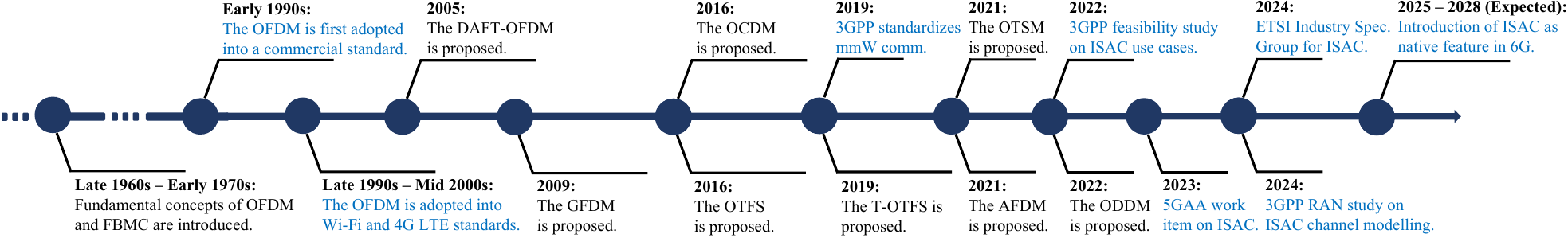}
\vspace{-3.5ex}
\caption{Timeline highlighting the invention of various waveforms, and key dates in standardization.}
\label{fig:waveform_timeline}
\vspace{-3ex}
\end{figure}

In this section, we briefly discuss various waveforms which, as illustrated in Fig. \ref{fig:waveform_map} and put into a chronological context in Fig. \ref{fig:waveform_timeline}, are related to the aforementioned \ac{OFDM}, \ac{OTFS}, and \ac{AFDM}, and can also be potential candidates to support \ac{ISAC} in \ac{B5G}/\ac{6G} systems.
Due to space limitation, however, the discussion is resumed to a qualitative comparison, addressed in more detail in Sec. \ref{sec:waveforms}.

\subsubsection{Intermediate Chirp Domain Waveforms} 
A few chirp-domain waveforms also exist, which in commonality with \ac{OFDM} and \ac{AFDM}, aim at orthogonalizing delay and Doppler shift indices.
Such waveforms, which include \ac{OCDM} \cite{Ouyang_TC16} and \ac{DAFT}-\ac{OFDM} \cite{Erseghe_TC05}, can be seen as special cases of \ac{AFDM}, with non-ideal and simplified chirp frequencies $c_1$ and $c_2$ \cite{Bemani_TWC23}, naturally exhibiting equal or worse performances depending on the doubly-dispersive channel profile.

\subsubsection{Enhanced Delay-Doppler Waveforms}
Various methods adopt the novel delay-Doppler signal representation of \ac{OTFS}, and have proposed enhanced delay-Doppler domain waveforms.
Examples are the \ac{T-OTFS} \cite{Surabhi_TWC19}, which maximizes the asymptotic diversity of \ac{OTFS} via a phase-rotating precoder; \ac{OTSM} modulation \cite{Thaj_TWC21}, which seeks to reduce implementation complexity by leveraging a new type of domain transform; and \ac{ODDM} \cite{Lin_TWC22}, which improves upon \ac{OTFS} via an optimized pulse-shaping filter that creates feasible pulses that are orthogonal with respect to the delay-Doppler plane resolution.

\subsubsection{Filter Bank-based (Pulse-Shaping) Waveforms}
Finally, various multicarrier techniques leverage optimized pulse-shaping filter banks, such as \ac{FBMC} \cite{Farhang_SPM11} and \ac{GFDM} \cite{Michailow_TC14}, which improve the \ac{OOB} emissions, spectral efficiency, \ac{ISI}, and \ac{ICI} problems of \ac{OFDM} via robust adaptation of the subcarriers and modulation pulses with respect to the doubly-dispersive channel statistics.

\newpage

\section{Integrated Sensing and Communications (\ac{ISAC}) using \\ Next-Generation Waveforms}
\label{sec:radar_sensing}

In \ac{ISAC}, wireless ``sensing'' refers to the ability to harness the rich information about the surrounding environment inherently embedded in radio signals affected by channel conditions.
\Ac{ISAC} techniques leveraging a single waveform\footnote{This is to distinguish from \ac{RCC} techniques \cite{Zhang_CST22}, which ``share'' the totally available resource (spectrum, antennas, etc.) to allocate and deploy both of the dedicated radar and communications signals.} can be classified as either \textit{a)} communication-centric designs -- which aim to enable the sensing functionality via primary communications waveforms (OFDM, OTFS, AFDM, etc.) by extracting environment/target information from the received signal; \textit{b)} radar-centric designs -- which aim to enable the communications functionality via primary radar waveforms (pulse, FMCW, etc.) by embedding information to the modulated signal, or \textit{c} dual-functional joint designs -- which aim to enable both functionalities via a novelly designed waveform not specific to either of the \ac{ISAC} functionalities.
In this article, we investigate the communication-centric designs in doubly-dispersive channels, \textit{i.e.,} enabling the sensing functionality with the aforementioned OFDM, OTFS, and AFDM communications waveforms.

Then, drawing a parallel with well-established \textit{radar} technologies \cite{RichardsBook2005} two distinct types of sensing problems can be identified, namely: 1) the \textbf{\acf{DP}}, which relates to resolving the number of unique scattering points of interest (targets) from the background clutter, and 2) the \textbf{\acf{EP}}, which refers to extracting parameters such as range, velocity, and bearing associated with the targets.
Such sensing scenarios can be classified as monostatic, where the sensing receiver is colocated with the transmitter, or bistatic, where the sensing receiver is spatially remote from the transmitter\footnote{The bistatic scenario can be further generalized into a multistatic scenario with multiple distributed transmitters and receivers, for several applications and use cases \cite{Rou_TWC24}.
This scenario is beyond the scope of this article but is an important setting to be addressed in future work, which can enable the full potential of \ac{ISAC}}.

\begin{figure}[t]
\centering
\includegraphics[width=0.85\textwidth]{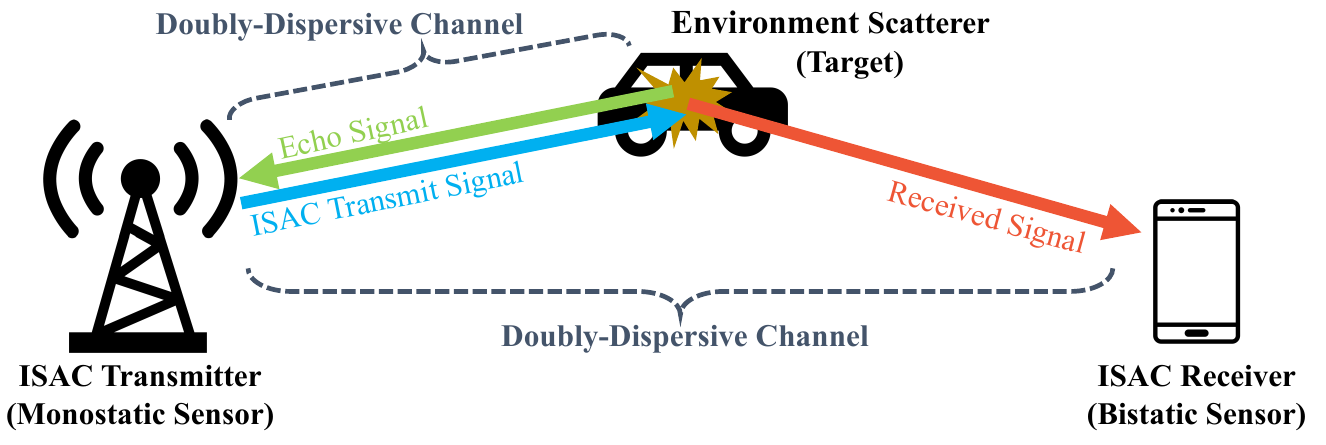}
\vspace{-1ex}
\caption{An illustration of the communications and sensing scenarios involved in an ISAC system.}
\label{fig:ISAC_scenario}
\vspace{-2.5ex}
\end{figure}

A exemplary \ac{ISAC} scenario between transmitter and receiver is illustrated in Fig. \ref{fig:ISAC_scenario} with a single environment scatterer, where the three sub-scenarios corresponding to the \ac{ISAC} functionalities and sensing locations can be observed.
Namely, there exist both communications and bistatic sensing subsystems between the \ac{ISAC} transmitter and the \ac{ISAC} receiver over the same doubly-dispersive channel of the scattering environment, and only the monostatic sensing subsystem for the \ac{ISAC} transmitter.

The radar parameters, namely the effective range $r$ [m] and relative linear velocity $v$ [m/s] of the target are directly correspondant to the path delay $\tau$ and Doppler shift $\nu$ of the target reflected signal, \vspace{-0.5ex}
\begin{equation}
{tau \triangleq \frac{r}{c},} ~~~~~\text{and}~~~~~ \nu \triangleq \frac{2v f_\mathrm{c}}{c},
\label{eq:radar_parameters} \vspace{-0.5ex}
\end{equation}
where $c$ is the speed of light and $f_\mathrm{c}$ is the carrier frequency of the transmitted signal.

For monostatic sensing, the effective range $r$ is twice the range of the target from the colocated transceiver (round-trip distance), whereas for bistatic scenarios, $r$ is the sum of the transmitter-to-target and receiver-to-target ranges (total propagation distance).
Similarly, for monostatic scenarios, the relative velocity $v$ is in the direction along a straight line between the target and the transceiver, whereas for bistatic scenarios, $v$ is in the direction along the bisecting line between the transmitter-to-target and receiver-to-target paths (bistatic velocity). 
In light of the relationship in eq. \eqref{eq:radar_parameters}, it is important to notice that the fundamental radar target parameters (range and velocity) are equivalent to the path dispersion parameters (delay and Doppler) of the common doubly-dispersive channel, as illustrated in Fig. \ref{fig:ISAC_scenario}.

Solutions to the \acp{DP} and \acp{EP} can be further classified into \textbf{three methods} based on their approaches: a) \textbf{correlation-based} methods, in which echo signals are filtered with the known transmit signal in order to yield the radar parameter estimates; b) \textbf{direct \ac{CSI}-based} methods, where radar parameters are extracted \emph{a posteriori} from the known channel matrix; and c) \textbf{indirect \ac{CSI}-based} methods, where radar parameters estimation and channel estimation are performed jointly, by leveraging the known structure of the doubly-dispersive channel of eq. \eqref{eq:IO_matrix}. 
Each of these approaches is discussed further in the sequel.

In the scope of this article, we address only the radar parameter estimation of the \ac{SISO} model, as the \ac{MIMO} extension can be trivially derived by leveraging the models in eq. \eqref{eq:received_signal_MIMO} and eq. \eqref{eq:beamform_channel} with the corresponding \ac{MIMO} radar processing techniques \cite{Li_Book08}.
This enables the estimation of bearing, \textit{i.e.,} the angular direction of the scatterer targets, and in the case of \ac{2D} planar arrays, the azimuth and elevation angles of the target in \acs{3D} space. 

\subsection{Correlation-based Methods}
\label{subsec:correlation_method}

Classical radar systems based on chirps and impulsive waveforms \cite{RichardsBook2005} are typical examples of the correlation-based method, as the received echo signals are processed by a correlation (matched filter) with a known probing signal, to directly yield the target parameters.
Radar waveforms are, however, optimized to exhibit correlation properties that can achieve high resolution in the delay and Doppler domains, which is generally not possible to do with communication waveforms without sacrificing communication objectives (e.g., rate, efficiency, latency, etc.).

The fundamental resolution of the correlation-based method can be analyzed through the well-known \ac{2D} ambiguity function of a waveform in delay-Doppler domain\footnote{In the case for \ac{MIMO}, the dimension of the ambiguity function is extended to include the angular domain.}, which is given by
\begin{equation}
A(\tau, \nu) \triangleq \int_{-\infty}^{+\infty} s(t)  \cdot s^*(t - \tau) \cdot e^{j2\pi\nu t} ~{\rm d}t,
\end{equation}
where $A(\tau, \nu)$ is the \ac{2D} ambiguity function parametrized by the delay and Doppler shift values, $s(t)$ is the transmit signal, and $(\cdot)^*$ denotes the complex conjugation operation.

By inspecting the ambiguity function behaviors of the \ac{OFDM}, \ac{OTFS}, and \ac{AFDM} waveforms \cite{Zhu_Arxiv23}, the different delay-Doppler resolution and beamlobe behaviors of the waveforms can be observed.
Namely, it is found that that \ac{OFDM} shows high resolution in the delay domain, but not in Doppler, whereas \ac{OTFS} and \ac{AFDM} show moderate resolution in both domains simultaneously, with \ac{AFDM} exhibiting an adjustable mainlobe width with a trade-off in the two domains by leveraging the chirp frequency parameters $c_1$ and $c_2$.
Therefore naturally, correlation-based estimators of low complexity, employing \ac{OFDM}, \ac{OTFS}, and \ac{AFDM} \cite{Kumari_TVT18,Raviteja_RadarConf19, Zhu_Arxiv23}, have been proposed, but were found to be still fundamentally dependent on the resolution of the ambiguity function, promoting subsequent development of parameter estimation methods based on the higher resolution of the inherent delay-Doppler modulation grid.

Furthermore, it is important to notice that correlation-based methods inherently require the full knowledge of the transmit signal.
Therefore, in consideration of the \ac{ISAC} regime, its application is only feasible for monostatic sensing scenarios in which the transmit signal is fully known, or for bistatic scenarios by leveraging transmission frames with a known sequence, such as a pilot or preamble.

\vspace{-1.5ex}

\subsection{Direct \ac{CSI}-based Methods}
\label{subsec:directCSI_method}
These methods operate under the assumption that channel estimation (\ac{CSI} acquisition) has already been performed at the \ac{ISAC} receiver such that the effective channel matrix is available numerically, \textit{i.e.,} as complex taps of the channel matrix, and aim to extract radar parameters \emph{a posteriori} by exploiting the deterministic structure of the doubly-dispersive channel as described in Sec. \ref{sec:waveforms}.
Several sensing algorithms based on this approach have been proposed.
Assuming that \ac{CSI} is obtained in the time-frequency domain as per eq. \eqref{eq:TVTF}, the resulting estimation problem on $\tau_p$ and $\nu_p$ is referred to as a multidimensional harmonic retrieval problem, to which many well-known super-resolution solutions exist, such as \ac{MUSIC} and \ac{ESPRIT}, in addition to enhanced tensor-based algorithms in the case of \ac{MIMO} scenarios with increased dimensionality, as discussed in \cite{Zhang_CST22}.
The approach is known to achieve high resolution and accuracy but typically requires a large number of continuously obtained observation samples, which together with the fundamental dependence on the channel estimation performance, constitute a prominent challenge of such methods, since channel estimation errors may propagate to the radar parameter estimates. 

Setting these issues aside, with a sufficiently large number of observations and appropriate filtering, the effective channel matrices of the waveforms as given in eqs. \eqref{eq:eff_ofdm}, \eqref{eq:eff_otfs}, and \eqref{eq:eff_afdm} may be obtained via compressive sensing and other matrix reconstruction algorithms.
In such cases, the delay-Doppler orthogonality of the \ac{OTFS} and \ac{AFDM} effective channels, in addition to the injective mapping between each integer delay-Doppler pair and the position of the shifted-diagonal as discussed in Sec. \ref{sec:waveforms}, enable efficient extraction of the target parameters from the channel matrix element positions.
However, in the presence of fractional Doppler shifts, the resulting interference in the Doppler domain illustrated in Fig. \ref{fig:effChan_Frac} can significantly deteriorate such ``position-based" methods in terms of Doppler resolution.

Since the direct \ac{CSI}-based method inherently assumes the full knowledge of the effective channel matrix, which requires a dedicated channel estimation step involving a known transmit signal, this is also only possible for monostatic sensing scenarios or for bistatic scenarios with a dedicated pilot.

\subsection{Indirect \ac{CSI}-based Methods (Integrated Channel Estimation)}
\label{subsec:indirectCSI_method}

In cases where the channel matrix is not available but its structure is known, an estimation problem can be formulated to jointly estimate the channel inherently incorporating the radar parameters, yielding an integrated channel and target parameter estimation.
Such scenarios are expected at \ac{ISAC} receivers where the information of the transmit signal is available but channel estimation has not been performed, or at bistatic \ac{ISAC} receivers with unknown components of the transmit signal (\textit{i.e.,} information symbols).

When the transmit signal is fully known, the following minimization problem can be solved
\begin{equation}
\underset{\tau_p,\nu_p,h_p \,\forall p}{\mathrm{argmin}} ~ \mathcal{L} \big( {\mathbf{y}} - {\tilde{\mathbf{G}}}(\tau_p,\nu_p,h_p; \,\forall p)\!\cdot\!{\mathbf{x}} \big),
\label{eq:3P_minimization}
\end{equation} 
where $\mathbf{y} \triangleq \mathbf{G}\mathbf{s} + \mathbf{w} \in \mathbb{C}^{N \times 1}$ is the {demodulated} received signal {where $\mathbf{G}$ is the effective channel corresponding to the specific waveform,} $\tilde{\mathbf{G}}(\tau_p,\nu_p,h_p; \,\forall p) \in \mathbb{C}^{N \times N}$ is the estimated {effective} channel parametrized by the $3P$ parameters $\tau_p,\nu_p,h_p$ for $p \in \{1,\cdots\!,P\}$, and $\mathbf{s} \in \mathbb{C}^{N \times 1}$ is the transmit {symbol vector}, and $\mathcal{L}(\cdot)$ is an arbitrary objective function, \textit{i.e.,} the $L_2$ norm.

In the doubly-dispersive case, the parametrized channel $\tilde{\mathbf{H}}$ is given by eq. \eqref{eq:IO_matrix}, or pre-processed via leveraging the effective channel models of specific waveforms described, e.g., by eqs. \eqref{eq:eff_ofdm}, \eqref{eq:eff_otfs}, or \eqref{eq:eff_afdm} for \ac{OFDM}, \ac{OTFS} and \ac{AFDM}, respectively.

On the otherhand, when parts of the transmit signal is also unknown, a bilinear estimation is required in the form of a \ac{JCDE} problem on the system described by \vspace{-1ex}
\begin{equation}
\mathbf{y} = \tilde{\mathbf{G}}(\tau_p, \nu_p, h_p; \forall p) \!\cdot\! \tilde{\mathbf{x}} \in \mathbb{C}^{N \times 1}, \vspace{-1ex}
\end{equation}
where $\mathbf{y} \in \mathbb{C}^{N \times 1}$ is the demodulated received signal, $\tilde{\mathbf{G}}(\tau_p, \nu_p, h_p; \forall p) \in \mathbb{C}^{N \times N}$ is the estimated effective channel matrix parametrized by the $3P$ parameters, $\tilde{\mathbf{x}} \triangleq \in \mathbb{C}^{N \times N}$ is the estimated transmit symbol vector composed of known pilot symbols and unknown information symbols.

The joint estimation of the unknown data symbols and the unknown channel parametrized by the $3P$ radar parameters, is therefore an inherently equivalent problem to the joint target detection and data estimation, which can be solved via a plethora of methods as will be elaborated in the following section.

\vspace{-1.25ex}
\subsection{Estimation Methods of \ac{CSI}-based Methods}
\label{subsec:CSI_ISAC_Method}
\vspace{-0.5ex}

There are largely two approaches to the underlying $3P$-parameter estimation problem present in \ac{CSI} matrix-based sensing methods in Sections \ref{subsec:directCSI_method} and \ref{subsec:indirectCSI_method}: \textit{a)} on-grid search-based methods which yield values corresponding to the resolution of the discretized parameter grid, and \textit{b)} off-grid methods which directly yield a continuous estimate of the parameters.

On-grid methods \cite{Keskin_ICC21, Ranasinghe_ICASSP24} assume the domain of the path delay and Doppler shift to be discrete on a given grid resolution, typically of the discretely sampled signal and channel, and solve the optimization problem via \ac{ML}, \textit{i.e.,} a search over the discrete parameter solution space.
Typically, the parameter grid is initially limited to the sampling frequency, which are the integer values of the normalized delay and normalized digital Doppler frequency.
However, a common approach is to first obtain the on-grid ``coarse" integer estimates of the parameters, then perform a iterative search around the estimate with a virtually sub-divided grid.
Such sub-divisions can be refined to arbitrarily high resolutions, but trivally require a higher complexity as the estimation requires the optimization over the sub-divided grid is still discrete.

On the other hand, off-grid methods directly yield a continuous estimate of the radar parameters by leveraging a variety of techniques methods including message passing, Bayesian learning, convex optimization, harmonic retrieval, \textit{etc.,} \cite{Liu_ICASSP21, Wei_TWC22}.
Such methods are not restricted by the resolution of the sampling frequency or a discrete grid, and capable of directly estimating the fractional components of the parameters, but as they are not \ac{ML}, may be more prone to the error caused by the noise and interference, and may exhibit and algorithmic resolution limit (\textit{i.e.,} error-floor). 

Furthermore, the fractional components of the Doppler shift cause spectral leakage around the main peaks of the effective channel as described in Sec. \ref{sec:waveforms} and illustrated in Fig. \ref{fig:effChan_Frac3D}, which makes the effective channel less sparse and subject to interference between the channel components of different paths.
Depending on the specific algorithm utilized for parameter estimation, such sub-optimal structures of the effective channel without proper alleviation techniques can deteriorate the sensing performance.


\section{Key Performance Indicator (\ac{KPI})-centered Comparative Analysis}
\label{sec:comparative_analysis}

In view of all the above, we finally offer a qualitative comparison of \ac{OFDM}, \ac{OTFS} and \ac{AFDM} -- respectively representing the classic, \ac{SotA} and most-recent alternative -- \ac{ISAC}-friendly waveforms for \ac{B5G}/\ac{6G} systems.
To this end, we consider various relevant \acp{KPI} for communication and sensing, functions, both in terms of features and implementation aspects.
The \ac{GFDM} and \ac{FMCW} waveforms are also included for comparison, for communications and radar sensing performances respectively.

The result is given in Table \ref{tab:comp_table}, and while it is not possible to elaborate on all comparison points due to space limitations, we briefly elaborate on a few most important of the selected \acp{KPI}.
In particular, perhaps the most important communications \ac{KPI} in doubly-dispersive channels is the Doppler-shift robustness, \textit{i.e.,} the compatibility to high-mobility and \ac{EHF} conditions, which is only attained by the \ac{OTFS} and \ac{AFDM} waveforms due to the inherent delay-Doppler domain orthogonality.
On this aspect, it is noteworthy that \ac{OTFS} achieves full diversity only in finite \ac{SNR} regimes converging to first order asymptotic diversity \cite{Surabhi_TWC19}, while \ac{AFDM} provides guaranteed full diversity generally \cite{Bemani_TWC23} and is also known to be the only Doppler-robust waveform to achieve full diversity also in \ac{MIMO} scenarios.
On the other hand, due to their fundamental roots on \ac{OFDM}, both \ac{AFDM} and \ac{GFDM} also suffer from higher \ac{PAPR}, whereas \ac{OTFS} enjoys a low \ac{PAPR} due to the \ac{DFT}-based spreading of the symbol powers in the time-frequency domain.
This advantage is closely linked to implementation cost and hardware stability, especially in relation to the \ac{PA} and \ac{RF} component efficiency which becomes more prominent in the massive \ac{MIMO} scenarios.

Another important point to consider is the computational complexity, which is linked to various signal processing procedures such as modulation and channel estimation.
In this regard, while both \ac{OTFS} and \ac{AFDM} can be interpreted as modified precoding schemes for \ac{OFDM} transmitters, such that core \ac{OFDM} modulators can be reused, the \ac{1D} \ac{AFDM} modulator exhibits a higher efficiency than the \ac{2D} \ac{OTFS} modulator.
This reduced dimension of the waveform also shows a similar advantage for channel estimation, in terms of both computational complexity and the required piloting overhead.

In terms of target sensing performance, \ac{OTFS} and \ac{AFDM} exhibit a significant improvement in the Doppler-domain ambiguity over \ac{OFDM}, but such methods are restricted as they are not optimized for correlation as with \ac{FMCW} waveforms.
Therefore, super-resolution methods and on-grid estimation methods on the discrete delay-Doppler domain of the waveforms are leveraged, which can achieve extremely high resolutions compared to those of \ac{FMCW}, often used in automotive radar, given sufficient parameterization such as the carrier frequency and symbol period.

The above-described properties in both communications and sensing performances must be satisfied and be coherent, in order for a waveform to be considered a strong candidate for \ac{ISAC} in \ac{B5G}/\ac{6G}.
Clearly, as observed from the color scaling of Table \ref{tab:comp_table}, \ac{OTFS} and \ac{AFDM} are the most promising candidates satisfying most of the \ac{ISAC} criteria, with some trade-offs between the two waveforms in terms of complexity, power and spectral efficiency, which is to be further addressed in a future work.

\begin{table}[t]
\centering
\caption{A comparative table of various waveforms and their \ac{ISAC} \acp{KPI}, with qualitative measures: high, medium, low. 
The color of each cell corresponds to the relative performance measure, ranging from {\color{teagreen} \bf green} denoting an attractive performance, {\color{flavescent} \bf yellow}, to {\color{indianred} \bf red} denoting less performant.
\vspace{-0.25ex}
}
\begin{tabular}{|c|c||l||c|c|c|c|}
\hline\hline
\multirow{2}{*}{} & 
\multirow{2}{*}{} & 
\multirow{2}{*}{\textbf{Key Performance Indicator}} & 
\multicolumn{4}{c|}{\textbf{Waveform}} \\ 
\cline{4-7} &  & &
\textbf{\acs{OFDM}} & 
\textbf{\acs{OTFS}} & 
\textbf{\acs{AFDM}} & 
\textbf{\acs{GFDM}}
\\ \hline \hline
%
\multirow{13}{*}{\rotatebox[origin=c]{90}{Communications}}
&
\multirow{7}{*}{\rotatebox[origin=c]{90}{Performance}}
& 
Waveform Domain & 
Time-frequency &
Delay-Doppler &
Chirp &
Filter-bank \\ \cline{3-7} & & \multicolumn{4}{c|}{} \\[-2.25ex] \cline{3-7}\\[-2.65ex]
& &
Delay Robustness &
\good High &
\good High &
\good High &
\good High \\ \cline{3-7}\\[-2.65ex]
& &
Doppler Robustness &
\bad Low &
\good High &
\good High &
\medium Medium \\ \cline{3-7}\\[-2.65ex]
& &
Peak-to-Average Power Ratio &
\bad High &
\good Low &
\bad High &
\medium Medium \\ \cline{3-7}\\[-2.65ex]
& &
Diversity in \ac{TV} Channels &
\bad Low &
\medium Medium &
\good High &
\medium Medium \\ \cline{3-7}\\[-2.65ex]
& &
Frame Guard (CP) Overhead &
\bad High &
\good Low &
\medium Medium &
\medium Medium \\ \cline{3-7}\\[-2.65ex]
& &
Pilot Guard Overhead &
\medium Medium &
\bad High &
\good Low &
\medium Medium \\
\cline{2-7}\\[-2.25ex]
\cline{2-7}\\[-2.65ex]

& \multirow{6}{*}{\rotatebox[origin=c]{90}{Implementation}}
& & & & & \\[-2.71ex] 
& &
Modulation Complexity &
\good Low &
\bad High &
\medium Medium &
\bad High \\ \cline{3-7}\\[-2.65ex]
& &
\ac{OFDM} Compatibility &
\good High &
\good High &
\good High &
\good High\\ \cline{3-7}\\[-2.65ex]
& &
\Ac{PA} Strain &
\bad High &
\good Low &
\bad High & 
\medium Medium \\ \cline{3-7}\\[-2.65ex]
& &
\ac{MIMO} Scalability &
\good High &
\medium Medium &
\good High &
\medium Medium \\ \cline{3-7}\\[-2.65ex]
& &
\ac{EHF} Feasibility &
\bad Low &
\good High &
\good High &
\medium Medium \\ \cline{3-7}\\[-2.65ex]
& &
Full-Duplex Potential &
\bad Low &
\medium Medium &
\good High &
\medium Medium \\
\hline\hline
\end{tabular}

\vspace{1.5ex}

\begin{tabular}{|c|c||l||c|c|c|c|}
\hline\hline
\multirow{2}{*}{} &
\multirow{2}{*}{} & 
\textbf{Key Performance Indicator} &
\textbf{\acs{OFDM}} & 
\textbf{\acs{OTFS}} & 
\textbf{\acs{AFDM}} & 
\textbf{\acs{FMCW}}
\\[-2.75ex]
& & &
\hphantom{Time-frequency} & 
\hphantom{Delay-Doppler} & 
\hphantom{Chirp} & 
\hphantom{Filter-bank}\\ \hline \hline
\multirow{9}{*}{\rotatebox[origin=c]{90}{Target Sensing}}
&
\multirow{5}{*}{\rotatebox[origin=c]{90}{Performance}}
&
Delay Ambiguity &
\good Low &  
\medium Medium & 
\medium \textit{Variable} & 
\good Low \\ \cline{3-7}\\[-2.65ex]
& &
Doppler Ambiguity &
\bad High &  
\medium Medium & 
\medium \textit{Variable} & 
\good Low \\ \cline{3-7} \\[-2.65ex]
& &
Peak-to-Sidelobe Ratio &
\medium Medium &  
\good Low & 
\medium \textit{Variable}  &
\good Low \\  \cline{3-7} \\[-2.65ex]
& &
Max. Unambiguous Range &
\medium Medium &  
\medium Medium & 
\medium Medium & 
\good High \\ \cline{3-7} \\[-2.65ex]
& &
Max. Unambiguous Velocity &
\medium Medium &  
\medium Medium & 
\medium Medium & 
\good High \\ \cline{2-7} \\[-2.25ex] \cline{2-7} \\[-2.65ex]

%
&
\multirow{4}{*}{\rotatebox[origin=c]{90}{Implem.}}
& \multicolumn{4}{c|}{} \\[-2.72ex] \\[-2.65ex]
& & 
Implementation Cost &
\good Low &
\medium Medium &
\medium Medium & 
\good Low \\ \cline{3-7} \\[-2.65ex]
& & 
Engineering Complexity &
\good Low &
\medium Medium &
\medium Medium & 
\good Low \\ \cline{3-7} \\[-2.65ex]
& & 
MIMO Array Extendibility &
\good High &
\medium Medium &
\good High &
\good High\\ \cline{3-7} \\[-2.65ex]
& & 
ISAC Feasibility &
\medium Medium &
\good High  &
\good High &
\medium Medium \\ 
\hline \hline
\end{tabular}
\label{tab:comp_table}
\vspace{-3ex}
\end{table}

\section{Future Works}
\label{sec:future_work}

It can be seen that while \ac{OTFS} and \ac{AFDM} are the most promising candidates to enable high-performance \ac{ISAC} for next-generation wireless networks, there are still many important topics to be addressed in order to enable the incorporation of such techniques into future standards.
Indeed, any given row of Table \ref{tab:comp_table} -- such as the \ac{PAPR} of \ac{AFDM} or the \ac{CSI} estimation complexity of \ac{OTFS} -- can be a subject of optimization and development.
The authors hope that this article helps the \ac{ISAC} research community with fundamental insights, techniques, and future direction to promote the development of high-performance \ac{ISAC} in doubly-dispersive environments for next-generation wireless networks.
Furthermore, in light of the consolidated fundamentals and the identification of potentials and challenges of the doubly-dispersive environment and the promising waveforms, it is also important to investigate the implications towards the consequent practical implementations and design.



\newpage

\section*{Author Biography}

\vspace{-15ex}

\begin{IEEEbiography}[{\includegraphics[width=1in,height=1.25in,clip,keepaspectratio]{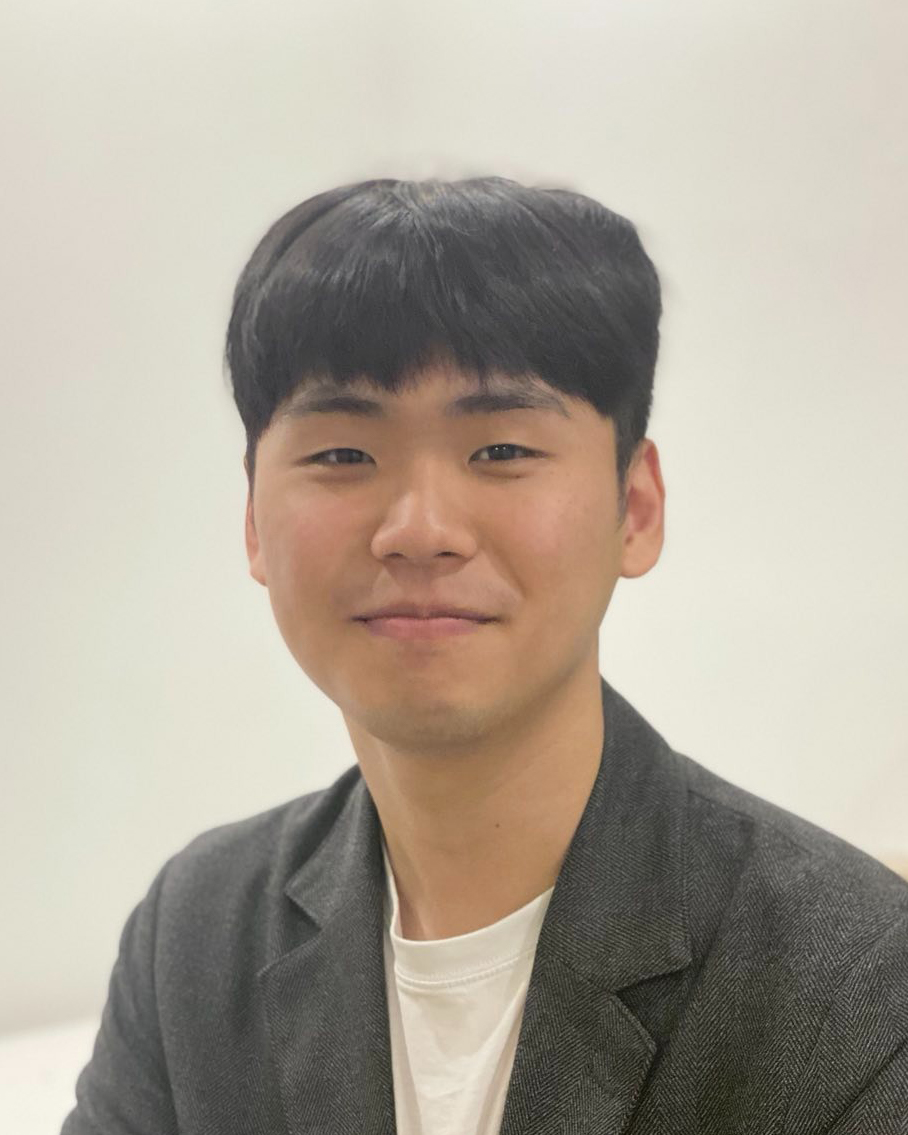}}]{Hyeon Seok Rou} [\textit{hrou@constructor.university}] (Graduate Student Member, IEEE) has received the Ph.D. degree in Electrical Engineering in 2024 from Constructor University, Bremen, Germany, and the B.Sc. degree in Electrical and Computer Engineering (ECE) in 2021 from Jacobs University Bremen, Germany. He has received the Korea Institute of Science and Technology Europe Research Scholarship Award from The Korean Scientists and Engineers Association in the FRG (Verein Koreanischer Naturwissenschaftler und Ingenieure in der BRD e.V.) in 2022, and he was a visiting researcher at the Intelligent Communications Lab, Korea Advanced Institute of Science and Technology (KAIST) in 2023. His research interests include integrated sensing and communications (ISAC), signal processing in doubly-dispersive channels, high-mobility communication systems, multi-dimensional modulation, next-generation metasurfaces, B5G/6G V2X communication technologies, and quantum computing.
\end{IEEEbiography}

\vspace{-10ex}

\begin{IEEEbiography}[{\includegraphics[width=1in,height=1.25in,clip,keepaspectratio]{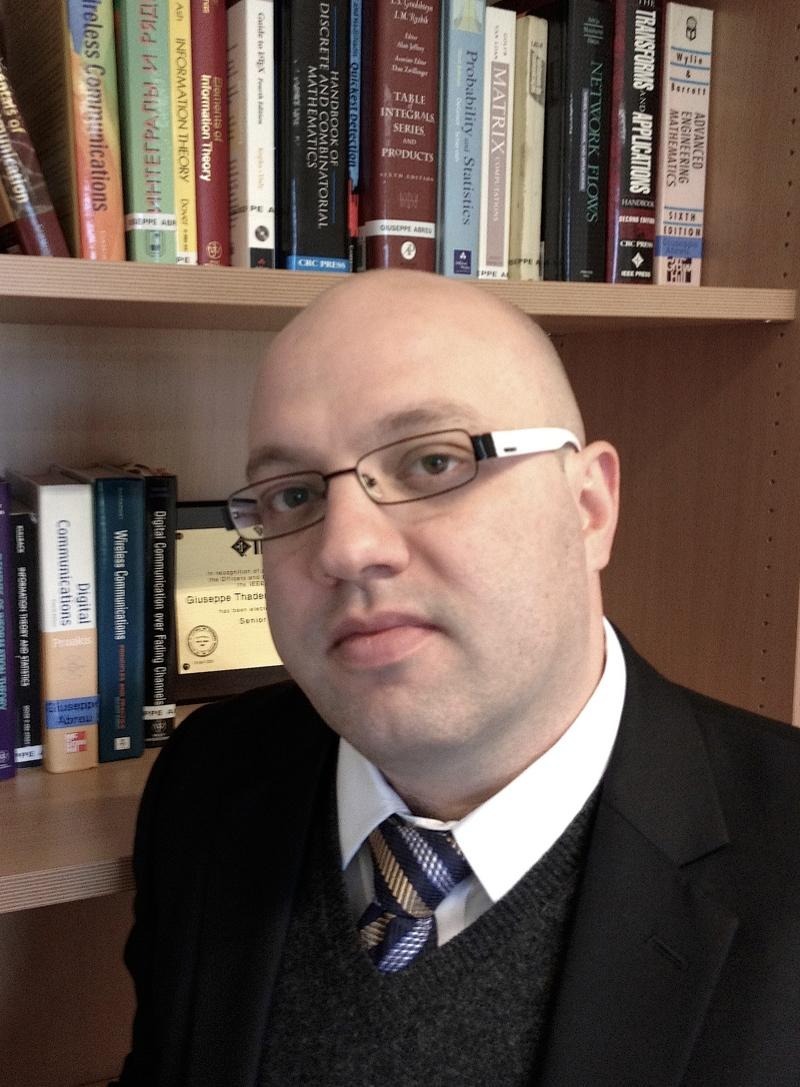}}]{Giuseppe Thadeu Freitas de Abreu} [\textit{gabreu@constructor.university}] (Senior Member, IEEE) is a Full Professor of Electrical Engineering at Constructor University, Bremen, Germany. His research interests include communications theory, estimation theory, statistical modeling, wireless localization, cognitive radio, wireless security, MIMO systems, ultrawideband and millimeter wave communications, full-duplex and cognitive radio, compressive sensing, energy harvesting networks, random networks, connected vehicles networks, and many other topics. He has served as an editor for various IEEE Transactions, and currently serves as an editor to the IEEE Signal Processing Letters and the IEEE Communications Letters.
\end{IEEEbiography}

\vspace{-10ex}

\begin{IEEEbiography}[{\includegraphics[width=1in,height=1.25in,clip,keepaspectratio]{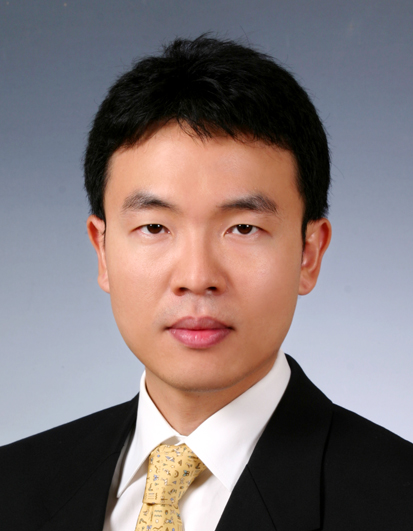}}]{Junil Choi} [\textit{junil@kaist.ac.kr}] (Senior Member, IEEE) received the Ph.D. degree from Purdue University, USA, in 2015, and is currently an Associate Professor School of Electrical Engineering, Korea Advanced Institute of Science and Technology (KAIST), South Korea. He has received four IEEE journal paper awards from the IEEE Communications, Signal Processing, and Vehicular Technology Societies, and is currently serving as an Associate Editor of various IEEE journals and letters. His research interests include the design and analysis of massive MIMO, mmWave communication systems, distributed reception, and vehicular communication systems.
\end{IEEEbiography}

\vspace{-10ex}

\begin{IEEEbiography}[{\includegraphics[width=1in,height=1.25in,clip,keepaspectratio]{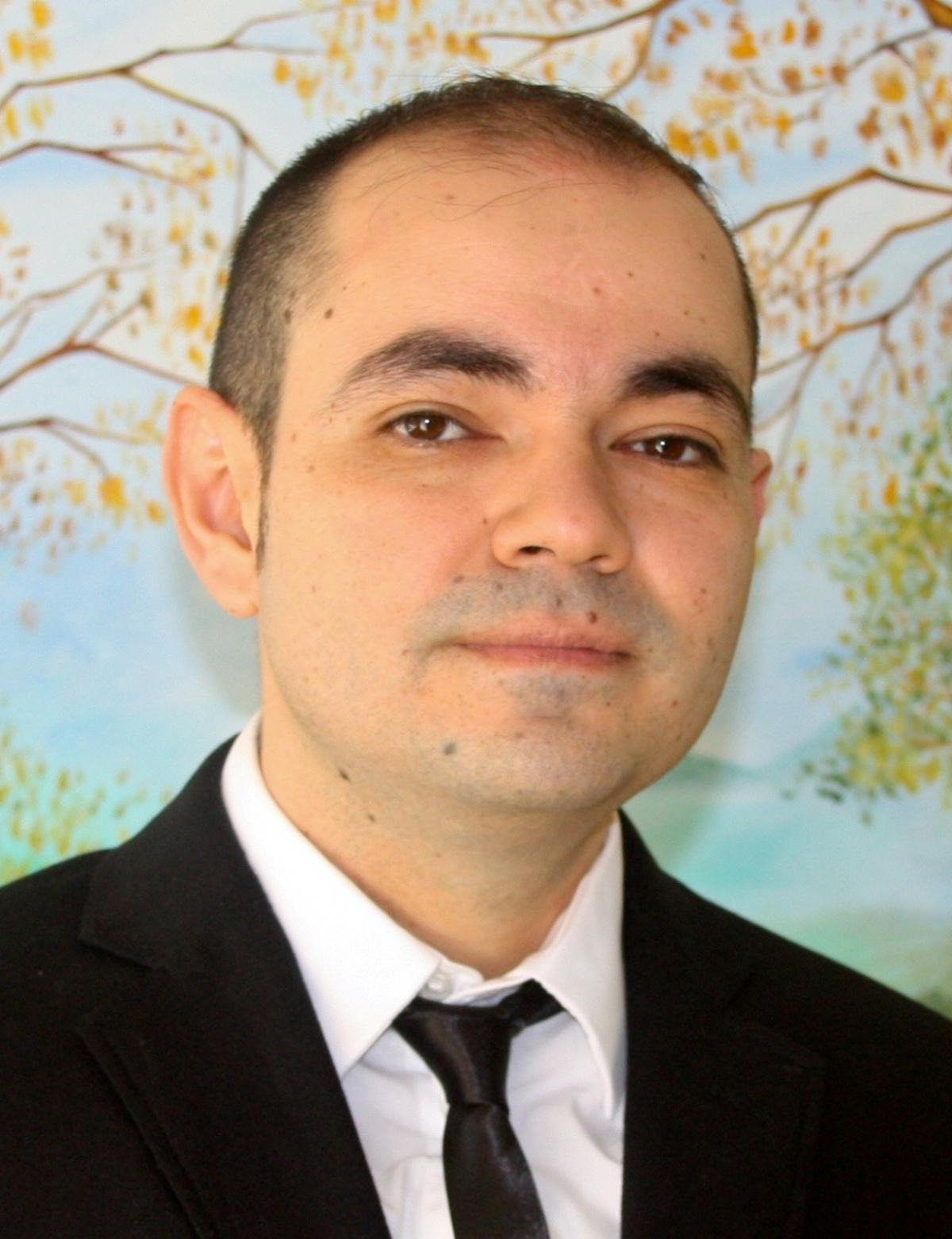}}]{David Gonz\'{a}lez G.} [\textit{david.gonzalez.g@ieee.org}] (Senior Member, IEEE) has a master in Mobile Communications and PhD in Signal Theory and Communications from the Universitat Politècnica de Catalunya, Spain. He has served as post-doctoral fellow in Aalto University, Finland (2014-2017). He also served as Research Engineer with Panasonic Research and Development Center, Germany. Since 2018, David is with Continental Automotive Technologies, Germany, where he conducts and manages research projects focused on diverse aspects of vehicular communications (V2X), integrated sensing and communications (ISAC), and automotive applications for 5G-Advanced and 6G. David participates as delegate in 3GPP RAN1, 5GAA, and ETSI ISG ISAC.
    
\end{IEEEbiography}

\newpage

\begin{IEEEbiography}[{\includegraphics[width=1in,height=1.25in,clip,keepaspectratio]{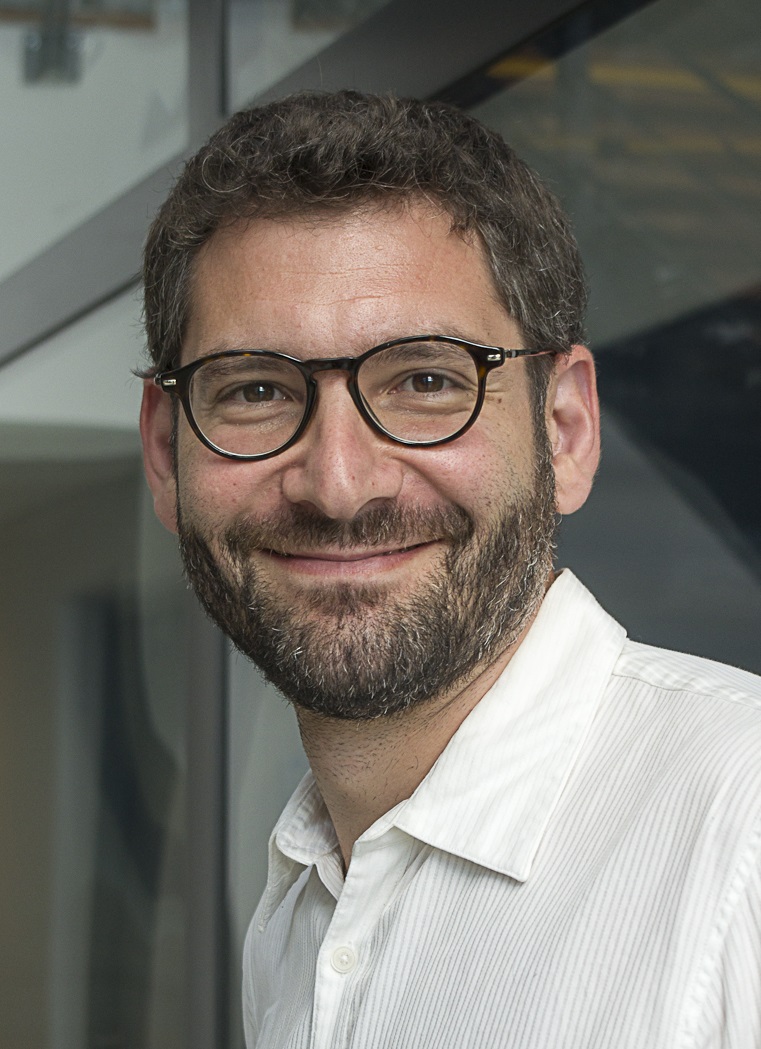}}]{Marios Kountouris} [\textit{kountour@eurecom.fr}] (IEEE Fellow) is a Professor at the Communication Systems Department, EURECOM, France, and a Distinguished Researcher at the Department of Computer Science and Artificial Intelligence, University of Granada, Spain. He has held positions at Centrale-Supélec, France, University of Texas at Austin, USA, Huawei Paris Research Center, France, and Yonsei University, South Korea. He is the recipient of a European Research Council Consolidator Grant on goal-oriented semantic communication. He has received several awards and distinctions, including the 2022 Blondel Medal, the 2016 IEEE ComSoc CTTC Early Achievement Award, and the 2013 IEEE ComSoc EMEA Outstanding Young Researcher Award.
\end{IEEEbiography}

\vspace{-10ex}

\begin{IEEEbiography}[{\includegraphics[width=1in,height=1.25in,clip,keepaspectratio]{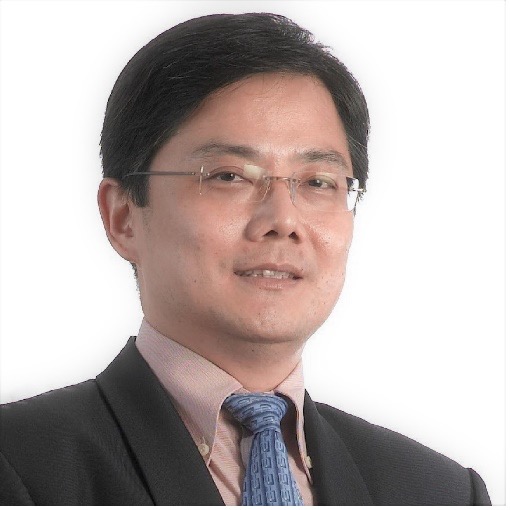}}]{Yong Liang Guan} [\textit{eylguan@ntu.edu.sg}] (Senior Member, IEEE) obtained his PhD degree from the Imperial College London, and Bachelor of Engineering degree with first class honours from the National University of Singapore.  He is now an Associate Vice President of the Nanyang Technological University (NTU), Singapore, and a Professor of Communication Engineering at the School of EEE in NTU.  His research interests broadly include coding and signal processing for communication and data storage systems.  He has published a monograph, 2 books and 550 journal and conference papers. He has 50 filed patents and 7 granted patents (some of which were licensed to NXP, Continental). 
\end{IEEEbiography}

\vspace{-10ex}

\begin{IEEEbiography}[{\includegraphics[width=1in,height=1.25in,clip,keepaspectratio]{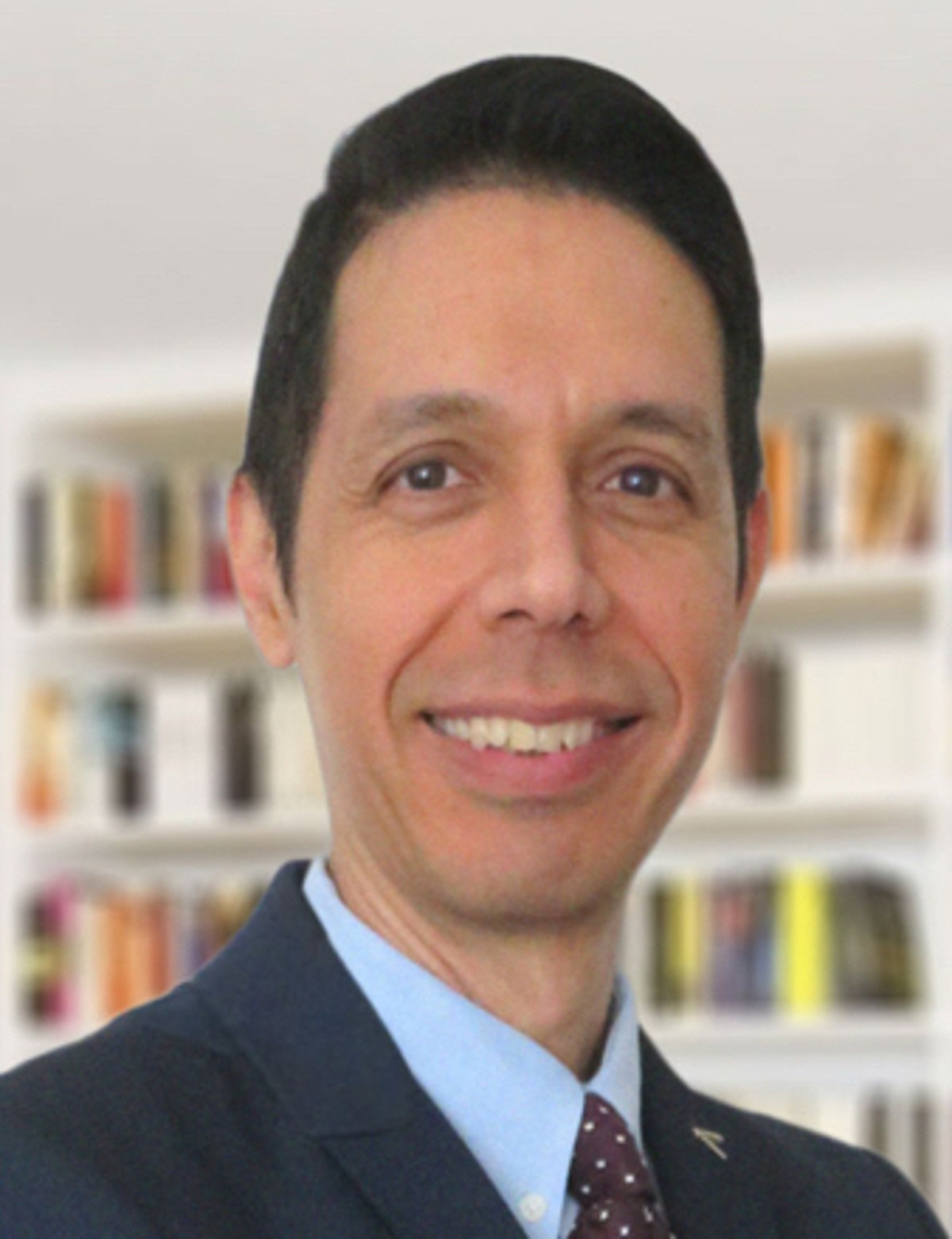}}]{Osvaldo Gonsa} [\textit{osvaldo.gonsa@continental.com}] received his PhD degree in electrical and computer engineering from Yokohama National University, Japan, in 1999, and the M.B.A. degree from the Kempten School of Business, Germany, in 2012. He has led the 5G-Advanced/6G research and standardization efforts, and founded the Wireless Communications Technologies Group at the Software and Central Technologies organization of Continental Automotive Technologies, Frankfurt, Germany. He has numerous patents and has co-authored several books, in addition to numerous scientific papers. He serves as a member for the GSMA Advisory Board for automotive and is the main representative of Continental at the 5GAA association.
\end{IEEEbiography}

\vfill

\end{document}